\documentclass[journal]{IEEEtran}
\ifCLASSINFOpdf
\else
\fi
\usepackage[comma,numbers,square,sort&compress]{natbib}
\usepackage{algorithm} 
\usepackage{algorithmic} 
\usepackage{arydshln}
\usepackage{multirow} 
\usepackage{xcolor}
\usepackage{setspace}
\usepackage{amsmath}
\newtheorem{theorem}{\textbf{Theorem}}
\newtheorem{lemma}{\textbf{Lemma}}
\newtheorem{example}{\textbf{Example}}
\newtheorem{corollary}{\textbf{Corollary}}
\newtheorem{remark}{\textbf{Remark}}
\newtheorem{definition}{\textbf{Definition}}

\newtheorem{proposition}{\textbf{Proposition}}

\newenvironment{proof}{{\noindent{\bf \emph{Proof:}}}\quad}{\hfill $\square$\par}

\usepackage{multirow}
\usepackage{url}
\usepackage{subfigure}
\usepackage{fancyhdr}
\usepackage{amsmath}
\usepackage{multirow}
\usepackage{amssymb }
\usepackage{color}
\usepackage{graphics} 
\usepackage{graphicx}
\usepackage{geometry}

\renewcommand{\baselinestretch}{1.00} \normalsize

\begin{document}
	%
	%
	%
	%
	
	\title{\LARGE \bf
			Functional Observability, Structural Functional Observability and Optimal Sensor Placement
	}
	
	%
	%
	
	\author{Yuan Zhang, Tyrone Fernando, and Mohamed Darouach
		\thanks{Y. Zhang is with the School of Automation, Beijing Institute of Technology, Beijing, China (email: zhangyuan14@bit.edu.cn). T. Fernando is with the Department of Electrical Electronic and Computer Engineering, University of Western Australia,  Crawley, WA 6009, Australia (email: tyrone.fernando@uwa.edu.au). M. Darouach is with the Centre de Recherche en Automatique de
			Nancy, IUT de Longwy, Universit\'{e} de Lorraine, 54400 Cosnes et Romain,
			France (email:	modar@pt.lu).} 
			\thanks{ The work of Dr. Zhang is supported by the National Natural Science Foundation of China under Grant 62373059. The work of Professor Darouach is supported by Robert and Maude Gledden, Senior Visiting Fellowship, Institute of Advanced Studies and by the Forrest fellowship, UWA, Australia.}} 
	\maketitle

	\begin{abstract}In this paper, new characterizations for functional observability, functional detectability, and structural functional observability (SFO) are developed, and based on them, the related optimal sensor placement problems are investigated.  A novel concept of {\emph{modal functional observability}} coinciding with the notion of modal observability is proposed. This notion introduces necessary and sufficient conditions for functional observability and detectability in a unified way without resorting to system observability decomposition, and facilitates the design of a functionally observable/detectable system. Afterwards, SFO is redefined rigorously from a generic perspective, contrarily to the definition of structural observability. A complete graph-theoretic characterization for SFO is proposed. Based on these results, the problems of selecting the minimal sensors from a prior set to achieve functional observability and SFO are shown to be NP-hard. Nevertheless, supermodular set functions are established, leading to greedy heuristics that can find approximation solutions to these problems with provable guarantees in polynomial time. A closed-form solution along with a constructive procedure is also given for the unconstrained case on systems with diagonalizable state matrices. Notably, our results also yield a polynomial-time verifiable case for structural target controllability, a problem that may be hard otherwise.

	\end{abstract}
	

	\section{INTRODUCTION}

 	Large-scale complex systems are prevalent in the real world, encompassing various domains such as engineering, biology, and social networks \cite{Y.Y.2011Controllability,liu2013observability,gao2014target}. In these systems, the acquisition of accurate and timely information through sensing and estimation is crucial for effective feedback control \cite{Fixed_Mode,Stefano2012Stability}. However, as the dynamical networks grow in size and complexity, it often becomes impractical or even infeasible to deploy sensors for every individual state variable. For many real-world problems, it is neither necessary nor desirable to estimate the entire state vector of a high-dimensional system \cite{liu2013observability,montanari2022functional}.   Instead, the focus often lies on specific subsets or linear functions of state variables that are of particular interest or relevance \cite{niazi2020average}. The design of functional observers to estimate linear functions of states addresses this need~\cite{darouach2000existence,fernando2010functional}. An analogous scenario is to control a target subset of state variables, rather than the whole state variables in complex systems \cite{gao2014target}. 

The idea of designing functional observers could date back to the pioneering work of Luenberger \cite{luenberger1966observers}. Since then, a significant amount of research has been devoted to the functional observer design problems, ranging from linear systems \cite{darouach2000existence,fernando2010functional,rotella2011minimal,jennings2011existence,rotella2015note} to nonlinear systems \cite{trinh2006partial}. Particularly, Darouach  \cite{darouach2000existence} gave a complete algebraic condition for the existence of a minimum order functional observer. Fernando et al. \cite{fernando2010functional}, Rotella et al. \cite{rotella2011minimal}, and {Darouach and Fernando} \cite{Darouach2023FunctionalDA,functional2022Mohamed,asymptotic2022Mohamed} addressed the design of reduced-order functional observers when the aforementioned condition was not satisfied by augmenting the functional of states to be estimated.  { Functional observers have been applicable to various practical scenarios, such as disturbance observation of servo control systems \cite{kim2013reduced,su2018further}, cyberattack detection and epidemic spreading estimation \cite{montanari2022functional}, average state monitoring of clustered network systems \cite{niazi2020average,niazi2023clustering}, and load frequency control of real-world power systems \cite{pham2015load,alhelou2019decentralized}, just to name a few. } 

It was very recently that the concept of functional observability was proposed by Fernando et al. \cite{fernando2010functional}. Functional observability encompasses the ability to design an observer with arbitrary poles to estimate a linear function of states asymptotically from system inputs and outputs. Notably, it has been discovered that functional observability can be alternatively defined as the capability to infer the function of states solely from system external outputs \cite{fernando2010functional2,jennings2011existence,functional2022Mohamed}.  A closely related concept to functional observability is functional detectability, which was first proposed by Fernando et al. \cite{fernando2010functional} and further developed by {Darouach and Fernando} \cite{Darouach2023FunctionalDA}. Functional detectability is a weaker constraint than functional observability, which enables the existence of an observer with stable poles to track a functional of states asymptotically. These generalizations of concepts of state observability and detectability have sparked interest in developing algebraic criteria that characterize functional observability and detectability, independent of the design of functional observers. {In this regard, similar to state observability, existing conditions for functional observability fall into two classes, those involving products of system matrices (e.g., system observability matrix), and those involving eigenvalues and eigenspaces of system state matrices, resembling the PBH test for state observability \cite{jennings2011existence,rotella2015note}. The advantage of the latter class of conditions lies in its capability to deconstruct the roles of system eigenvalues/modes, thus providing unified conditions for functional observability and detectability, akin to the PBH test for state observability and detectability \cite{jennings2011existence,Darouach2023FunctionalDA}.} Unfortunately, as we can see and also pointed out in \cite{functional2022Mohamed,asymptotic2022Mohamed} recently, the PBH-like conditions are generally only necessary. {A complete eigenvalue and eigenspace based characterization would be desirable. As we shall show, such a characterization not only offers a unifying view for conditions of functional observability and detectability, but also provides a deconstructed way to design system measurement structure for achieving functional observability/detectability (see Section \ref{sec-III}).}




Furthermore, the concept of {\emph{structural functional observability (SFO)}} has recently been proposed by Montanari et al. \cite{montanari2022functional}. This novel notion defines SFO as the existence of a functionally observable realization for a linear system, considering its known zero-nonzero structure. Similar to other structural properties like structural controllability \cite{C.T.1974Structural}, SFO leverages the inherent system structure, specifically the network topology of large-scale systems. By avoiding dependence on precise numerical values of system parameters, SFO enables scalable combinatorial algorithms for functional observability verification and sensor placement optimizations, getting rid of parameter uncertainty and rounding-off errors. The notable feature of SFO offers a complementary perspective that is particularly valuable for the analysis and design of large-scale systems. However, as demonstrated later, the SFO has not been defined rigorously, and its graph-theoretic criterion given in \cite{montanari2022functional} relies heavily on a PBH-like rank condition for functional observability, making its sufficiency questionable.   

%
%

In this paper, we develop new characterizations for functional observability, functional detectability, and SFO {that are valid for general systems}, and leverage them to the associated system design problem, more precisely, optimally deploying sensors to achieve functional observability/detectability. The contributions of this work are fourfold:


First, we propose a novel concept of {\emph{modal functional observability}}, which generalizes the notion of modal observability and naturally introduces new necessary and sufficient conditions for functional observability and detectability from an eigenspace perspective. The obtained conditions directly facilitate the measurement structure design of a functional observable/detectable system, as they only concern the property of each individual mode and are independent of the system canonical observability decomposition. 


Second, we justify that the notion of SFO cannot be solely defined as the existence of a functionally observable realization and redefine it rigorously from a generic property perspective. As a result, SFO exhibits some dramatically different properties from structural observability, one of which is that adding additional links among states or from states to sensors may destroy SFO.

Third, we give a complete graph-theoretic characterization for SFO, utilizing the concept of {\emph{maximum independent walk family}}. This characterization provides valuable insights into how the system structure influences SFO.  A by-product finding is that, verifying the structural target controllability of $n-1$ state variables is achievable in polynomial time, where $n$ is the system state dimension. This gives a positive answer to a special case of the long-standing structural target controllability verification problem \cite{murota1990note,czeizler2018structural}.

Fourthly, based on the obtained criteria, we prove that the problems of selecting the minimal sensors from a prior set to achieve functional observability and SFO are NP-hard. Nevertheless, we establish supermodular functions that lead to greedy heuristics to find near-optimal solutions for these problems with approximation guarantees. We also give a closed-form solution and a constructive procedure to design an output matrix with the minimum number of rows to achieve functional observability for diagonalizable systems. This is based on modal functional observability and the real Jordan normal form. {By considering only unstable modes, these results carry over for functional detectability.}

The rest is organized as follows. Section \ref{pre-sec} introduces some preliminaries. Section \ref{section_prob} presents the definitions of functional observability and detectability, and Section \ref{sec-III} is devoted to modal functional observability and new criteria for functional observability and detectability. SFO is redefined in Section \ref{Sec-IV}. A complete graph-theoretic characterization is given for SFO in Section \ref{Sec-V}. The minimal sensor placement problems are addressed in Section \ref{Sec-VI}. The last section concludes this paper.

\section{Preliminaries} \label{pre-sec}
This section introduces notational conventions and some preliminaries in graph theory, structured system theory, and submodular set function.

\subsection{Notations}
Define $[n]\doteq \{1,2,...,n\}$ for an integer $n\ge 1$.  Denote the set of eigenvalues of a square matrix $M$ by ${\rm eig}(M)$. For a number $\lambda\in {\mathbb C}$, ${\rm Re}(\lambda)$ denotes the real part of $\lambda$.  Let ${\mathbb C}^-$ and ${\mathbb C}^+_0$ be the open left and the closed right halfplanes of the complex plane, respectively.  Let ${\bf col}\{X_i|_{i=1}^n\}$ be the composite matrix stacked by $X_1,...,X_n$, with $X_i$ being the $i$th row block. When $n$ is small, ${\bf col}\{X_1,...,X_n\}$ is also denoted by $[X_1;...;X_n]$.  Similarly, ${\bf diag}\{X_i|_{i=1}^n\}$ denotes a block-diagonal matrix whose diagonal blocks are $X_1,...,X_n$.  $I_n$ denotes the $n$ dimensional identify matrix, where the subscript $n$ may be omitted if inferred from the context. Given an $m\times n$ matrix $M$ and a set $S\subseteq [m]$, ${M}_{S}$ denotes the sub-matrix of $M$ formed by {\emph{rows}} indexed by $S$. The space spanned by {\emph{rows}} of a matrix $M$ is given by ${\cal R}(M)$. The number of rows of a matrix $M$ is denoted by ${\rm row}(M)$.  The dimension of a subspace is denoted by ${\rm dim}(\cdot)$.

\subsection{Graph theory} \label{basic_notion}

Let ${\cal G}=(V,E)$ be a directed graph (digraph) with vertex set $V$ and edge set $E\subseteq V\times V$. A subgraph ${\cal G}_s=(V_s,E_s)$ of ${\cal G}$ is a graph such that $V_s\subseteq V$ and $E_s\subseteq E$, and is called a subgraph induced by $V_s$ if $E_s=(V_s\times V_s) \cap E$. We say ${\cal G}_s$ {\emph{covers}} $V_s'$, if $V_s'\subseteq V_s$.  A path $P$ from $i_1$ to $i_k$ in ${\cal G}$ is a sequence of edges $(i_1,i_2)$, $(i_2,i_3)$,...,$(i_{k-1},i_k)$ with $(i_j,i_{j+1})\in E$, $j=1,...,k-1$, which is also denoted by $P=(i_1,i_2,...,i_k)$. If $i_1,i_2,...,i_k$ are distinct vertices, then $P$ is called a simple path. If $i_1$ and $i_k$ are the only repeated vertices in $P$, then $P$ is a {\emph{cycle}}. The length of a path is the edges it contains.

\subsection{ Structured system theory} \label{preliminary-structured-system}
Consider a linear-time invariant system described by	
\begin{subequations}  \label{pre-system}
	\begin{align}
		\dot x(t)&=Ax(t)+Bu(t), \label{pre-state-eq}\\
		y(t)&=Cx(t), \label{pre-lumped-output}
	\end{align}
\end{subequations}
where $x(t)\in {\mathbb R}^{n}$ is the state vector, $u(t)\in{\mathbb R}^{m} $ is the input vector, $y(t)\in {\mathbb R}^{p}$ is the output vector, $A\in {\mathbb R}^{n\times n}$, $B\in {\mathbb R}^{n\times m}$, and $C\in {\mathbb R}^{p\times n}$. System (\ref{pre-system}) is also denoted by the pair $(A,C)$.

A structured matrix is a matrix with entries that are either fixed zeros or undetermined, capable of taking arbitrary values independently. Denote the set of $n_1\times n_2$ structured matrices by $\{0,*\}^{n_1\times n_2}$, where $*$ represents an undetermined entry, and $0$ a fixed zero entry.
For a structured matrix $\bar M\in \{0,*\}^{n_1\times n_2}$, a realization is obtained by assigning some specific values to the $*$ entries (termed free entries), and the set of all realizations is given by ${\cal S}(\bar M)=\{M\in {\mathbb R}^{n_1\times n_2}: M_{ij}=0 \ {\rm if} \ \bar M_{ij}=0 \}$.


Given a structured pair $(\bar A,\bar C)$, with $\bar A\in \{0,*\}^{n\times n}$ and $\bar C\in \{0,*\}^{p\times n}$, $(A,C)$ is called a realization of $(\bar A, \bar C)$ if $A\in {\cal S}(\bar A)$ and $C\in {\cal S}(\bar C)$. $(\bar A,\bar C)$ is said to be structurally observable, if there is a observable realization $(A,C)$ of $(\bar A, \bar C)$ \cite{C.T.1974Structural}.


For a structured pair $(\bar A, \bar C)$, let ${\cal G}(\bar A, \bar C)=(X\cup Y, E_{XX}\cup E_{XY})$ be its {\emph{system digraph}}, with state vertices $X=\{x_1,...,x_n\}$, output vertices $Y=\{y_1,...,y_p\}$, and edges $E_{XX}=\{(x_i,x_j): \bar A_{ji}\ne 0\}$, $E_{XY}=\{(x_i,y_j): \bar C_{ji}\ne 0\}$. A state vertex $x_i\in X$ is said to be {\emph{output-reachable}}, if there is a path from $x_i$ to some output vertex $y_j\in Y$ in ${\cal G}(\bar A, \bar C)$. Vertex $x_i\in X$ has a self-loop if $(x_i,x_i)\in E_{XX}$.
An {\emph{output stem}} is a simple path from a state vertex to an output vertex in ${\cal G}(\bar A, \bar C)$.

{\begin{definition}
		A subset of state vertices $X_s\subseteq X$ is said to be {\emph{covered by a cactus configuration}} in ${\cal G}(\bar A, \bar C)$, if 1) every $x_i\in X_s$ is output-reachable in ${\cal G}(\bar A, \bar C)$, and 2) $X_s$ can be covered by a collection of vertex-disjoint cycles and output stems in ${\cal G}(\bar A, \bar C)$.
\end{definition}}

The following result is well-known in structured system theory.

\begin{lemma}\cite{generic,Ramos2022AnOO} \label{structural-observability-theorem}
	The pair $(\bar A, \bar C)$ is structurally observable, if and only if the whole state vertex set $X$ is {\emph{covered by a cactus configuration}} in ${\cal G}(\bar A, \bar C)$.
\end{lemma}

\subsection{Submodular set function}\label{preliminary-submodular}
Given a finite set $V=\{1,2,...,p\}$, a {\emph{set function}}: $f: 2^V\to {\mathbb R}$ assigns a scalar value to a subset of $V$.  A set function $f: 2^V\to {\mathbb R}$ is said to be {\emph{monotone decreasing}}, if for all $V_1\subseteq V_2\subseteq V$, it holds $f(V_1)\ge f(V_2)$; $f$ is {\emph{monotone increasing}} if $-f$ is monotone decreasing.  A set function $f: 2^V\to {\mathbb R}$ is said to be {\emph{submodular}}, if for all $V_1\subseteq V_2\subseteq V$ and all elements $s\in V$, $f(V_1\cup \{s\})-f(V_1)\ge f(V_2\cup \{s\})-f(V_2)$; $f$ is {\emph{supermodular}}, if $-f$ is submodular ({see \cite[Def. 1.2]{krause2014submodular}}). Roughly speaking, submodularity is such a property that adding an element to a smaller set gives a larger gain than adding one to a larger set, and supermodular is on the contrary.  For any $s\in V$, define the gain function $f_a: 2^{V\backslash\{a\}}\to {\mathbb R}$ as $f_a(S)=f(S\cup \{a\})-f(S)$. {By definition}, $f$ is submodular if $f_a$ is monotone decreasing for each $a\in V$.

Submodular set functions in combinatorial optimization play a similar role to convex functions in continuous optimization \cite{T2016On}. It is known that a simple greedy heuristic can find near-optimal solutions to minimization of monotone decreasing supermodular functions with provable guarantees \cite{wolsey1982analysis}.

\section{Functional observability and detectability} \label{section_prob}
Consider a linear-time invariant system described by	
\begin{subequations}  \label{lumped-system}
	\begin{align}
		\dot x(t)&=Ax(t)+Bu(t), \label{state-eq}\\
		y(t)&=Cx(t), \label{lumped-output} \\
		z(t)&=Fx(t),   \label{functional-info}
	\end{align}
\end{subequations}
where $x(t)\in {\mathbb R}^{n}$ is the state vector, $u(t)\in{\mathbb R}^{m} $ is the input vector, $y(t)\in {\mathbb R}^{p}$ is the output vector, and $z(t)\in {\mathbb R}^{r}$ is the functional of states to be estimated. Accordingly, matrices $A\in {\mathbb R}^{n\times n}$, $B\in {\mathbb R}^{n\times m}$, $C\in {\mathbb R}^{p\times n}$, and $F\in {\mathbb R}^{r\times n}$. Each nonzero row of $C$ corresponds to a sensor. System (\ref{lumped-system}) is also denoted by the triple $(A,C,F)$.\footnote{Since matrix $B$ does not affect the results of this paper, we ignore $B$ in the matrix triple representation of system (\ref{lumped-system}).} Throughout, we use $O(A,C)={\bf col}\{C,CA,...,CA^{n-1}\}$ to denote the observability matrix of $(A,C)$.


The notion of {\emph{functional observability}} of the triple $(A,C,F)$ was first proposed by Fernando et al. \cite{fernando2010functional}, which generalizes the conventional notion of observability. Functional observability can be defined in a variety of ways, either the existence of an observer with arbitrary poles that can track the functional state vector $z(t)$ asymptotically \cite{fernando2010functional}, or the ability to infer $z(t)$ from system external outputs and inputs. Those definitions are shown to be equivalent \cite{fernando2010functional2,jennings2011existence}. 





\begin{definition}[Functional observability, \cite{fernando2010functional2}]\label{functional-def} System (\ref{lumped-system}) (or the triple $(A,C,F)$) is said to be functionally observable, if for any initial state $x(0)$ and input $u(t)$, there exists a finite time $T$ such that the value of $Fx(0)$ can be uniquely determined from the outputs $y(t)$ and inputs $u(t)$, $0\le t \le T$.
\end{definition}

By the additivity of linear systems, $(A,C,F)$ is functionally observable, if and only if for any initial state $x(0)$ and the zero input $u(t)\equiv 0$, $y(t)=Cx(t)=0$, $\forall t\in [0,\infty)$ implies that $z(t)=Fx(t)= 0$ \cite{functional2022Mohamed}. It is easy to see that when $F=I_n$, functional observability collapses to the conventional observability. It is shown in \cite{fernando2010functional} that $(A,C,F)$ is functionally observable, if and only if there exists a matrix $F_0\in {\mathbb R}^{\bar r\times n}$ with ${\cal R}(F)\subseteq {\cal R}(F_0)$ satisfying
\begin{equation} \label{functional-observer-condition}\begin{array}{c}
		{\rm rank}[F_0A;CA;C;F_0]={\rm rank}[CA;C;F_0],\\
		{\rm rank}[sF_0-F_0A;CA;C]={\rm rank}[CA;C;F_0], \forall s\in {\mathbb C},\end{array}
\end{equation}where $r\le \bar r \le n$, such that an observer with order $\bar r$ and arbitrary poles can be constructed to estimate $z(t)$ asymptotically \cite{darouach2000existence}. {Recently, functional observability has been found to bear close relationship with differential privacy and the privacy protection of system initial values from being inferred by adversarial eavesdroppers in \cite{zhang2023observability,wang2023differential}. The underlying concept posits that the initial value privacy $z(0)=Fx(0)$ is protected if and only if $z(t)$ is not functionally observable.}

A related concept to functional observability is {\emph{functional detectability}} \cite{Darouach2023FunctionalDA}, formally defined as follows.

\begin{definition}[Functional detectability, \cite{Darouach2023FunctionalDA}] \label{def-functional-detect}
	System (\ref{lumped-system}) (or the triple $(A,C,F)$) is said to be functionally detectable, if for any initial state $x(0)$ and the zero input $u(t)\equiv 0$, we have $y(t)=Cx(t)=0$, $\forall t\in [0,\infty)$ implies that $\lim_{t\to \infty} z(t)=0$.
\end{definition}

It was pointed out in \cite{Darouach2023FunctionalDA} that a functional observer with stable poles exists that can track the functional state $z(t)$ asymptotically, if and only if $(A,C,F)$ is functionally detectable. Obviously, by taking $F=I_n$, functional detectability reduces to the conventional concept of detectability. While functional observability implies functional detectability, the reverse direction is not always true. Moreover, $(A,C,F)$ is functionally detectable {if and only if} there exists a matrix $F_0\in {\mathbb R}^{\bar r\times n}$ satisfying
\begin{equation}\label{functional-detectability-condition}\begin{array}{c}{\rm rank}[F_0A;CA;C;F_0]={\rm rank}[CA;C;F_0],\\
		{\rm rank}[sF_0-F_0A;CA;C]={\rm rank}[CA;C;F_0], \forall s\in {\mathbb C}_0^+,
	\end{array}
\end{equation}
where ${\cal R}(F)\subseteq {\cal R}(F_0)$ and $r\le \bar r \le n$. {However, since $\bar r$ is undetermined and there may be infinitely many $s$ to be verified in the second equation of (\ref{functional-detectability-condition}), finding an $F_0$ satisfying the above equations is a challenging task~\cite{fernando2010functional,Darouach2023FunctionalDA}.} 



The goal of this paper is to develop new characterizations for functional observability, functional detectability, and SFO (defined in the sequel), and then leverage the obtained results to the associated sensor placement problems, more precisely, deploying the minimal sensors to achieve (structural) functional observability. 

\section{Modal functional observability and new criteria for functional observability/detectability} \label{sec-III}

	This section introduces a novel concept of modal functional observability, leading to new necessary and sufficient conditions for functional observability and functional detectability.

	Suppose $A$ has $k$ distinct eigenvalues, and denote the $i$th one by $\lambda_i$, $i=1,...,k$.	Let the invertible matrix $T\in {\mathbb C}^{n\times n}$ transform $A$ to its Jordan normal form $J$, i.e., $T^{-1}AT=J$.
	Recall that the Jordan normal form $J$ is a block diagonal matrix. Assume that the $i$th diagonal block of $J$, given by $J_i$, is associated with the eigenvalue $\lambda_i$. Note that each $J_i$ may contain multiple Jordan blocks along its diagonal, each matrix of which has non-zero super-diagonal entries all equal to $1$, and main diagonal entries equal to $\lambda_i$. Let $d_i$ be the dimension of $J_i$, and {denote $t(i,j)$ as the $(d_1+\cdots+d_{i-1}+j)$th column of $T$. Define $T_i\doteq [t(i,1),t(i,2),\cdots, t(i,d_i)]$}. Accordingly, $T_i$ consists of $d_i$ generalized eigenvectors of $A$ associated with $\lambda_i$ \cite{Horn2013Matrix}.

	{
		Notice that $J$ and $T$ are typically complex-valued if some $\lambda_i$ is.
		The real Jordan normal form is introduced here (see \citep[Chap. 3.4]{Horn2013Matrix} for details). Its immediate relevance is peripheral to this section but proves invaluable in Section \ref{modal_sesnor_placement}, facilitating the design of real-valued output matrices based on the proposed modal functional observability. To this end, assume that $A\in {\mathbb R}^{n\times n}$ has $2k_c$ distinct non-real eigenvalues and $k_r$ distinct real eigenvalues ($k=2k_c+k_r$). There exists a real invertible matrix $\tilde T\in {\mathbb R}^{n\times n}$ such that $\tilde T^{-1} A \tilde T= J_r$ is a real block diagonal matrix with each block being a {\emph{real Jordan block}}. A real Jordan block is either identical to a complex Jordan block (corresponding to a real eigenvalue $\lambda_i$), or is a block matrix, consisting of $q$ $2\times 2$ blocks $D_i={\tiny \left[\begin{array}{cc}
				a_i & b_i \\
				-b_i & a_i
			\end{array} \right]}$ on the main block diagonal and $q-1$ blocks $I_2$ on the block superdiagonal, corresponding to a pair of complex conjugate eigenvalue $\lambda_i=a_i+b_i{\bf i}$ and $\lambda_i^*=a_i-b_i{\bf i}$ ($a_i,b_i\in {\mathbb R}$), where $q$ is the dimension of a Jordan block associated with $\lambda_i$ (notice that $\lambda_i$ may correspond to multiple Jordan blocks, and each one corresponds to a real Jordan block), ${\bf i}$ is the imaginary unit, and $(\cdot)^*$ takes the (entry-wisely) complex conjugate of a scalar or matrix.
		
	}

	

	Let $C_i=CT_i$ and  $F_i=FT_i$ for $i=1,...,k$. It is easy to see that ${\rm rank}O(A,C)={\rm rank}O(T^{-1}AT,CT)={\rm rank}O(J,CT)$.
	Building on $(J_i,C_i,F_i)$, we introduce the concept of modal functional observability.
	
	\begin{definition}[Modal functional observability] \label{modal_functional_ob}
		An eigenvalue $\lambda_i$ of $A$ is said to be modal functionally observable, if \begin{equation}\label{modal_functional}
			{\rm rank}\left[\begin{array}{c}
				O(J_i,C_i)\\
				F_i
			\end{array}\right]={\rm rank}O(J_i,C_i).  	\end{equation}
	\end{definition}
	
	
	The concept of {\emph{modal functional observability}} naturally generalizes the conventional modal observability \cite{A.M1989Measures}. As can be seen, when $F_i=I_{d_i}$ (or $F_i=T_i$), Definition \ref{modal_functional_ob} collapse to ${\rm rank}O(J_i,C_i)=d_i$, i.e., the observability of $(J_i,C_i)$. Since all other $J_j$'s have different eigenvalues from $J_i$, this is further equivalent to ${\rm rank}[J-\lambda_iI;CT]={\rm rank}[A-\lambda_iI;C]=n$, which is exactly the definition of modal observability of $\lambda_i$.

	\begin{lemma} \label{diagonal-modal}
		When the algebraic multiplicity of $\lambda_i$ equals its geometric multiplicity, $\lambda_i$ is modal functionally observable, if and only if
		\begin{equation}\label{PBH-modal}
			{\rm rank}  \left[\begin{array}{c}
				A-\lambda_i I_n \\
				C \\
				F
			\end{array}\right]={\rm rank}  \left[\begin{array}{c}
				A-\lambda_i I_n \\
				C
			\end{array}\right].
		\end{equation}
	\end{lemma}
	
	\begin{proof} In the above-mentioned case, {$J_i=\lambda_i I_{d_i}$. Consequently, $O(J_i,C_i)=[C_i;\lambda_iC_i;\cdots;\lambda_i^{d_i-1}C_i]$, yielding  ${\rm rank}O(J_i,C_i)={\rm rank}C_i$. Similarly, ${\rm rank}[O(J_i,C_i);F_i]={\rm rank}[C_i;F_i]$. As a result, (\ref{modal_functional}) is equivalent to ${\rm rank}[C_i;F_i]={\rm rank}C_i$.} Since different $J_j$ ($j=1,...,k$) have different eigenvalues, { $J_j-\lambda_iI_{d_i}$ is invertible if $j\ne i$. Due to the block-diagonal structure of $J-\lambda_iI$, it turns out that
			$${\rm rank}\left[\begin{array}{c}J-\lambda_iI\\
				CT\\
				FT\end{array}\right]=\sum\limits_{j\ne i}d_j+ {\rm rank}\left[ \begin{array}{c}J_i-\lambda_iI_{d_i}\\
				C_i\\
				F_i\end{array} \right],$$
			where the equality comes from elementary row transformations.  Similarly, we have 	
			$${\rm rank}\left[\begin{array}{c}J-\lambda_iI\\
				CT\end{array}\right]=\sum\limits_{j\ne i}d_j+ {\rm rank}\left[ \begin{array}{c}J_i-\lambda_iI_{d_i}\\
				C_i\end{array} \right].$$
			As $J=T^{-1}AT$ and $J_i=\lambda_i I_{d_i}$, it follows that ${\rm rank}[C_i;F_i]={\rm rank}C_i$, if and only if (\ref{PBH-modal}) holds}. 		
	\end{proof}
	
	The following two lemmas are required for further derivations. The first one reveals the relation between ${\rm rank}O(A,C)$ and $\sum_{i=1}^k {\rm rank}O(J_i,C_i)$. 
	
	\begin{lemma}\citep[Lem 8]{A.Ol2014Minimal} Let $P_i\in {\mathbb C}^{d_i\times d_i}$ be any invertible matrix, $i=1,...,k$. It holds that\footnote{The original \citep[Lem 8]{A.Ol2014Minimal} deals with the dual controllability matrix with a single input, where $J_i$ comprises a single Jordan block. It is trivial to extend that result to the observability matrix with multiple inputs when $J_i$ comprises multiple Jordan blocks.}\label{equal_subspace}
		$${\rm rank}O(A,C)=\sum\limits_{i=1}^k{\rm rank}O(P_iJ_iP_i^{-1},C_iP_i^{-1}).$$
	\end{lemma}

	\begin{lemma} \label{necessary-lemma}
		Given $M\in {\mathbb C}^{n\times n}$, $N\in {\mathbb C}^{r\times n}$, if $Ne^{Mt}v=0$ for $t\in [0,\infty)$ holds for {\emph{all}} vectors $v\in {\mathbb C}^n$, then $N=0$. Moreover, if ${\rm eig}(M)\subseteq {\mathbb C}_0^+$, then $\lim_{t\to \infty}Ne^{Mt}v=0$ for {\emph{all}} vectors $v\in {\mathbb C}^n$ implies that $N=0$.
	\end{lemma}
	{
		\begin{proof}
			Given a $v\in {\mathbb C}^n\backslash \{0^n\}$, $Ne^{Mt}v=0$ for $t\in [0,\infty)$ implies that $v$ is in the unobservable subspace of $(M,N)$ (i.e., the null space of $O(M,N)$) \citep[page 42]{H.Tr2012.Control}. Since this condition holds for all $v\in {\mathbb C}^n$,  it follows that the unobservable subspace of $(M,N)$ is ${\mathbb C}^n$. This only happens when $N=0$.
			Consider the second condition. For a given initial state $v\in {\mathbb C}^n$, the state response with zero input of system $(M,N)$ is $x(t)=e^{Mt}v$. Since ${\rm eig}(M)\subseteq {\mathbb C}_0^+$, we have $\lim_{t\to \infty} e^{Mt}v\ne 0$ for any $v\ne 0$. The condition $\lim_{t\to \infty}Ne^{Mt}v=0$ but $\lim_{t\to\infty} x(t)=\lim_{t\to \infty}e^{Mt}v\ne 0$ implies that $v$ is in the undetectable subspace of $(M,N)$ \citep[page 117]{H.Tr2012.Control}. Since this holds for all $v\in {\mathbb C}^n$ and ${\rm eig}(M)\subseteq {\mathbb C}_0^+$, it follows that the undetectable subspace of $(M,N)$ is ${\mathbb C}^n$. This only happens when $N=0$.
		\end{proof}
	}
	
	To reveal the connection between functional observability/detectability and modal functional observability, some canonical observability decompositions are needed. Let $(A,B,C,F)$ in (\ref{lumped-system}) be transformed to its observable canonical form by the invertible matrix $P\in {\mathbb R}^{n\times n}$ as follows
	\begin{equation}\label{observability-decomposition} \begin{array}{c} \left[\begin{array}{c}
				\dot {\tilde x}_o(t) \\
				\dot {\tilde x}_{\bar o}(t)
			\end{array}\right]=\underbrace{\left[\begin{array}{cc}
					\tilde A_{o} & 0 \\
					\tilde A_{21} & \tilde A_{\bar o}
				\end{array}\right]}_{PAP^{-1}}\left[\begin{array}{c}
				\tilde x_o(t) \\
				\tilde x_{\bar o}(t)
			\end{array}\right]+\underbrace{\left[\begin{array}{c}
					\tilde B_o \\
					\tilde B_{\bar o}
				\end{array}\right]}_{PB}u(t),\\
			y(t)=\underbrace{\left[\begin{array}{cc}
					\tilde C_o & 0
				\end{array}\right]}_{CP^{-1}}\left[\begin{array}{c}
				\tilde x_o(t) \\
				\tilde x_{\bar o}(t)
			\end{array}\right], z(t)=\underbrace{\left[\begin{array}{cc}
					\tilde F_o & \tilde F_{\bar o}
				\end{array}\right]}_{FP^{-1}}\left[\begin{array}{c}
				\tilde x_o(t) \\
				\tilde x_{\bar o}(t)
			\end{array}\right],\end{array}\end{equation}
	where $\tilde x_o(t)\in {\mathbb R}^l$ and $\tilde x_{\bar o}(t)\in {\mathbb R}^{n-l}$ represent the states of observable and unobservable subsystems of system (\ref{lumped-system}) \cite{H.Tr2012.Control}. Accordingly, $(\tilde A_o,\tilde C_o)$ is observable. Moreover, introduce an $(n-l)\times (n-l)$ invertible $Q$ such that
	\begin{equation} \label{small-decomp}
		Q\tilde A_{\bar o}Q^{-1}=\left[\begin{array}{cc}
			\tilde A_{\bar o1} & 0 \\
			0 & \tilde A_{\bar o2}
		\end{array}\right],\tilde F_{\bar o}Q^{-1}=[\tilde F_{\bar o1}, \tilde F_{\bar o2}]\end{equation}where $\tilde A_{\bar o1}\in {\mathbb C}^{l_s\times l_s}$, $\tilde A_{\bar o2}\in {\mathbb C}^{l_u\times l_u}$ ($0\le l_s,l_u\le n-l$), ${\rm eig}(\tilde A_{\bar o1})\subseteq {\mathbb C}^-$, ${\rm eig}(\tilde A_{\bar o2})\subseteq {\mathbb C}^+_0$, and $\tilde F_{\bar o1}$ ($\tilde F_{\bar o2}$) has the same number of columns as $\tilde A_{\bar o1}$ ($\tilde A_{\bar o2}$). It is easy to see that such $Q$ always exists, for example, taking $Q$ to be the invertible matrix that transforms $\tilde A_{\bar o}$ into its (real) Jordan normal form.
	
	It follows from (\ref{observability-decomposition}) and (\ref{small-decomp}) that there exists an invertible coordinate transformation $\tilde x(t)=\tilde Px(t)={\tiny{\left[\begin{array}{cc}
				I_{l} & 0 \\
				0 & Q
			\end{array}\right]}}Px(t)$ transforming system (\ref{lumped-system}) into {\small{
			\begin{equation}\label{observability-decomposition-further} \begin{array}{c} \left[\begin{array}{c}
						\dot {\tilde x}_o(t) \\
						\dot {\tilde x}_{\bar o1}(t)\\
						\dot {\tilde x}_{\bar o2}(t)\\
					\end{array}\right]={\left[\begin{array}{ccc}
							\tilde A_{o} & 0 & 0 \\
							\tilde A_{211} & \tilde A_{\bar o1} & 0 \\
							\tilde A_{212} & 0 & \tilde A_{\bar o2}
						\end{array}\right]}\left[\begin{array}{c}
						\tilde x_o(t) \\
						\tilde x_{\bar o1}(t)\\
						\tilde x_{\bar o2}(t)
					\end{array}\right]+{\left[\begin{array}{c}
							\tilde B_o \\
							\tilde B_{\bar o1}\\
							\tilde B_{\bar o2}
						\end{array}\right]}u(t),\\
					y(t)={\left[\begin{array}{ccc}
							\tilde C_o & 0 & 0
						\end{array}\right]}\left[\begin{array}{c}
						\tilde x_o(t) \\
						\tilde x_{\bar o1}(t)\\
						\tilde x_{\bar o2}(t)\\
					\end{array}\right],\\ z(t)={\left[\begin{array}{ccc}
							\tilde F_o & \tilde F_{\bar o1} & \tilde F_{\bar o2}
						\end{array}\right]}\left[\begin{array}{c}
						\tilde x_o(t) \\
						\tilde x_{\bar o1}(t)\\
						\tilde x_{\bar o2}(t)\\
					\end{array}\right].\end{array}\end{equation}}}

	\begin{proposition}\label{decomposition-based-criterion} Given a triple $(A,C,F)$, the following statements hold:
		
		1) it is functionally detectable, if and only if $\tilde F_{\bar o2}=~0$.
		
		2) it is functionally observable, if and only if $\tilde F_{\bar o}=0$.
	\end{proposition}
	
	\begin{proof}Since the transformation $\tilde x(t)=\tilde Px(t)$ is invertible, it suffices to consider system (\ref{observability-decomposition-further}). We first prove statement 1). Necessity: consider the initial state $\tilde x(0)=[0_{l\times 1};0_{l_s\times 1};\tilde x_{\bar o2}(0)]$, where $\tilde x_{\bar o2}(0)\in {\mathbb C}^{l_u}$. It turns out that $\forall \tilde x_{\bar o2}(0)\in {\mathbb C}^{l_u}$ with the zero input $u(t)\equiv 0$, the system response $y(t)=0$ and $z(t)=\tilde F_{\bar o2}e^{\tilde A_{\bar o2}t}\tilde x_{\bar o2}(0)$. From Lemma \ref{necessary-lemma}, to make $\lim_{t\to \infty}z(t)=0$ $\forall \ \tilde x_{\bar o2}(0)\in {\mathbb C}^{l_u}$, it is necessary that $\tilde F_{\bar o2}=0$. Sufficiency: suppose an initial state $\tilde x(0)=[\tilde x_o(0);\tilde x_{\bar o1}(0);\tilde x_{\bar o2}(0)]$ corresponds to $y(t)=\tilde C_oe^{\tilde A_ot} \tilde x_o(0)=0$, $\forall t$, with the zero input $u(t)\equiv 0$. Since $(\tilde A_o, \tilde C_o)$ is observable, this is possible only if $\tilde x_o(0)=0$, and thus $\tilde x_o(t)=0, \forall t$. As a result, $\tilde x_{\bar o1}(t)=e^{\tilde A_{\bar o1}t}\tilde x_{\bar o1}(0)$, and it turns out that $\lim_{t\to \infty}\tilde x_{\bar o1}(t)=0$ as $\tilde A_{\bar o1}$ is stable. Therefore, $\lim_{t\to \infty}z(t)=\lim_{t\to \infty}\bar F_{\bar o1}\tilde x_{\bar o1}(t)=0$.
		
		Based on Lemma \ref{necessary-lemma}, statement 2) can be proved using a similar argument to statement 1). The details are thus omitted. 	
	\end{proof}

	{From Proposition \ref{decomposition-based-criterion}, it is not difficult to obtain the following Corollary \ref{fundamental-theorem}. The equivalence of item i) and ii) of Corollary \ref{fundamental-theorem} was originally proposed in \cite[Theo. 5]{jennings2011existence}. However, the PBH-like condition (\cite[Theo. 4]{jennings2011existence}) utilized therein, as demonstrated in Corollary \ref{functional-theorem-2}, proves insufficient for general systems. The equivalence of item ii) and item iii) of Corollary \ref{fundamental-theorem} was reported in \cite{rotella2015note}.  Herein, due to its pivotal role in subsequent derivations, we provide a complete and self-contained proof for this corollary.}
	
	
	\begin{corollary}\label{fundamental-theorem}
		The following statements are equivalent:
		\begin{itemize}
			\item[i)] The triple $(A,C,F)$ is functionally observable.
			\item[ii)] $${\rm rank}\left[\begin{array}{c}
				O(A,C)\\
				O(A,F)
			\end{array}\right]={\rm rank}O(A,C).$$
			\item[iii)] $${\rm rank}\left[\begin{array}{c}
				O(A,C)\\
				F
			\end{array}\right]={\rm rank}O(A,C).$$
		\end{itemize}
	\end{corollary}
	
	\begin{proof} If condition iii) holds, then ${\cal R}(F)\subseteq {\cal R}(O(A,C))$. From \citep[Sec. 3.4]{H.Tr2012.Control}, in the observable canonical form (\ref{observability-decomposition}), $P$ can be chosen such that its first $l$ rows span ${\cal R}(O(A,C))$ and its last $n-l$ rows can be arbitrary such that $P$ is non-singular.   As $F=[\tilde F_o, \tilde F_{\bar o}]P$, we have $\tilde F_{\bar o}=0$ (noting that $[\tilde F_o, \tilde F_{\bar o}]=FP^{-1}$ is unique), leading to the functional observability of $(A,C,F)$ by Proposition \ref{decomposition-based-criterion}. On the other hand, if $\tilde F_{\bar o}=0$, then $F=[\tilde F_o, 0]P$, implying that ${\cal R}(F)\subseteq {\cal R}(O(A,C))$, leading to condition iii). This proves condition i) $\Leftrightarrow$ condition iii). Next, since ${\cal R}(F)\subseteq {\cal R}(O(A,F))$, condition ii) $\Rightarrow$ condition iii) is obvious. If condition iii) holds, then there are $E_i\in {\mathbb R}^{r\times p}$, $i=0,...,n-1$, such that $F=\sum_{i=0}^{n-1} E_iCA^{i}$. By the Cayley-Hamilton theorem, we have ${\cal R}(O(A,F))\subseteq {\cal R}(O(A,C))$, implying condition iii) $\Rightarrow$ condition ii).
	\end{proof}


	\begin{remark}
		It should be noted that statement 2) of Proposition \ref{decomposition-based-criterion} was first reported in \cite{jennings2011existence}, which was derived from the existence condition (\ref{functional-observer-condition}) for reduced-order functional observers, and the proof of item ii) of Corollary \ref{fundamental-theorem} based on the Kalman decomposition given in (\ref{observability-decomposition}) was reported in \cite{fernando2010functional2}.  Here, we directly derive these results from Definition \ref{functional-def}.
	\end{remark}

The following theorem gives necessary and sufficient conditions for functional observability and functional detectability in terms of {\emph{modal functional observability}}. It exhibits that modal functional observability severs a similar role to modal observability w.r.t. observability and detectability.

\begin{theorem} \label{theorem-func-observe}
	Given a triple $(A,C,F)$, the following statements hold:
	
	1) $(A,C,F)$ is functionally observable, if and only if each eigenvalue $\lambda_i\in {\rm eig}(A)$ is modal functionally observable.
	
	2) $(A,C,F)$ is functionally detectable, if and only if each unstable eigenvalue $\lambda_i\in {\rm eig}(A)\cap {\mathbb C}_0^+$ is modal functionally observable.
\end{theorem}

\begin{proof}
	The proof is accomplished with a refined decomposition on $(J,CT,FT)$, recalling that $(J,CT,FT)$ is the Jordan normal form transformed from $(A,C,F)$ by the invertible coordination transformation $x'(t)\leftarrowtail T^{-1}x(t)$. For each block $(J_i,C_i,F_i)$, let us do its canonical observability decomposition by a $d_i\times d_i$ invertible matrix $P_i$ as
	\begin{equation}\label{sys-decomp-observability}
	\begin{array}{c}   \left[\begin{array}{ccc}
	P_i & 0 & 0 \\
	0 & I_{p} & 0 \\
	0 & 0 & I_r
	\end{array}\right] \left[\begin{array}{c}
	J_i \\
	C_i \\
	F_i
	\end{array}\right]P_i^{-1}=\left[\begin{array}{cc}
	\tilde A_{oi} & 0\\
	\tilde A_{21i} & \tilde A_{\bar oi} \\
	\tilde C_{oi} & 0  \\
	\tilde F_{oi} & \tilde F_{\bar oi}
	\end{array}\right],	\end{array}
	\end{equation}where $(\tilde A_{oi}, \tilde C_{oi})$ is observable, and the remaining matrices have compatible dimensions. Without losing any generality, assume that $\lambda_1,...,\lambda_{k_0-1}\in {\mathbb C}^-$, and $\lambda_{k_0},...,\lambda_k\in {\mathbb C}_0^+$, $1\le k_0\le k$. {{Let $Q$ be an $n\times n$ permutation matrix such that
{\small
$$\begin{array}{c}Q\left[\begin{array}{ccccc} 	\tilde A_{o1} & 0 &\cdots & 0 & 0\\
	\tilde A_{211} & \tilde A_{\bar o1}  &\cdots & 0 & 0\\
	0 & 0 & \ddots & 0 & 0 \\
	0 & 0 & \cdots & \tilde A_{ok} & 0\\
	0 & 0 & \cdots & \tilde A_{21k} & \tilde A_{\bar ok} \\  \end{array}\right]Q^{-1} \\=\left[\begin{array}{cccccc}
\tilde A_{o1} & \cdots & 0 & 0 & \cdots & 0 \\
\vdots & \ddots & \vdots & \vdots & \ddots & \vdots \\
0 & \cdots & \tilde A_{ok} & 0 & \cdots & 0 \\
\tilde A_{211}	& \cdots & 0 & \tilde A_{\bar o1} & \cdots & 0 \\
\vdots & \ddots & \vdots & \vdots & \ddots & \vdots \\
0 & \cdots & \tilde A_{211k} & 0 & \cdots & \tilde A_{\bar ok}
 \end{array}\right]. \end{array} $$}Then, the invertible coordination transformation $x''(t)\leftarrowtail Q{\bf diag}\{P_1,...,P_k\}T^{-1}x(t)$ transforms $(A,C,F)$ into}}
	\begin{equation}\label{deeper-decomposition}
	\left[\begin{array}{ccc|ccccc}
	\tilde A_{o1} & \cdots & 0 & 0 & \cdots & 0 & \cdots & 0 \\
	\vdots & \ddots & \vdots & \vdots & & \vdots &  & \vdots \\
	0 & \cdots & \tilde A_{ok} & 0  & \cdots & 0 & \cdots & 0 \\
	\hline
	\tilde A_{211}	& \cdots & 0 & \tilde A_{\bar o1} & \cdots & 0 & \cdots & 0 \\
	\vdots  & \ddots & \vdots & \vdots & \ddots & \vdots & \cdots & 0 \\
	0 & \cdots & 0 & 0 & \cdots & \tilde A_{\bar ok_0} & \cdots & 0 \\
	\vdots  & \ddots & \vdots & \vdots & \vdots & \vdots & \ddots & \vdots \\
	0 & \cdots & \tilde A_{211k} & 0 & \cdots & 0 & \cdots & \tilde A_{\bar ok} \\
	\hline
	\tilde C_{o1} & \cdots & \tilde C_{ok} & 0 & \cdots & 0 & \cdots & 0 \\
	\hline
	\tilde F_{o1} & \cdots & \tilde F_{ok} & \tilde F_{\bar o1} & \cdots & \tilde F_{\bar ok_0} & \cdots & \tilde F_{\bar ok}
	\end{array}\right].
	\end{equation}Denote $A_o={\bf diag}\{\tilde A_{o1},\cdots, \tilde A_{ok}\}$ and $C_o=[\tilde C_{o1},\cdots, \tilde C_{ok}]$. Notice that different $\tilde A_{oi}$ have different eigenvalues ($i=1,...,k$). By Lemma \ref{equal_subspace}, we have ${\rm rank}O(A_o,C_o)=\sum_{i=1}^kO(\tilde A_{oi},\tilde C_{oi})=\sum_{i=1}^k{\rm row}(\tilde A_{oi})={\rm row}(A_o)$, meaning that $(A_o,C_o)$ is observable. Additionally, notice that $\sum_{i=1}^kO(\tilde A_{oi},\tilde C_{oi})=\sum_{i=1}^kO(J_i,C_i)={\rm rank}O(A,C)$ (Lemma \ref{equal_subspace}), implying that ${\rm rank}O(A_o,C_o)={\rm rank}O(A,C)$. As a result, $(A_o,C_o)$ is exactly the observable part of $(A,C)$ in its observable decomposition. By Proposition \ref{decomposition-based-criterion}, $(A,C,F)$ is functionally observable, if and only if $\tilde F_{\bar oi}=0$, $\forall i\in [k]$. From the observability decomposition (\ref{sys-decomp-observability}), this is equivalent to that each eigenvalue of $A$ is modal functionally observable.
	
	Furthermore, notice that ${\bf diag}\{\tilde A_{\bar o1},\cdots, \tilde A_{\bar o(k_{0}-1)}\}$ has only stable eigenvalues, and ${\bf diag}\{\tilde A_{\bar ok_0},\cdots, \tilde A_{\bar ok}\}$ contains only unstable eigenvalues. From Proposition \ref{decomposition-based-criterion}, $(A,C,F)$ is functionally detectable if and only if $\tilde F_{\bar oi}=0$, for $i=k_0,...,k$. Using Corollary \ref{fundamental-theorem} on $(J_i,C_i,F_i)$ yields that $\bar F_{\bar oi}=0$, if and only if the eigenvalue $\lambda_i$ is modal functionally observable. This leads to the second statement.
\end{proof}


{To illustrate the system decompositions utilized in the proof of Theorem \ref{theorem-func-observe}, {as well as the modal functional observability based criterion, below we provide an example.}
\begin{example}\label{decomposition-example}
	Consider a triple $(A,C,F)$ as
	$$\begin{array}{c}
	A=\left[\begin{array}{ccccc}
		-1& 0& 0& 0& 1\\
		0& 1& 0& 0& 0\\
		0& 1& 1& 0& 0\\
		0& 0& 0& 0& 0\\
		0& 0& 0& 0&-1
	\end{array}\right],\begin{array}{c}
   C=[0, 1, 0, 1, 1]; \\
   F=[1,1,0,0,0].
   \end{array}
	\end{array}$$
There is a matrix $T$ (given as follows) that transforms $(A,C,F)$ to the Jordan normal form
$$\begin{array}{c}
	J=\left[\begin{array}{cc|cc|c}
	1&1&0&0&0\\
	0&1&0&0&0\\
	\hline
	0&0&-1&1&0\\
	0&0&0&-1&0\\
	\hline
	0&0&0&0&0
	\end{array}\right], \begin{array}{c}
T={\tiny \left[ \begin{array}{ccccc}
		 0&0&1&0&0\\
		0&1&0&0&0\\
		1&0&0&0&0\\
		0&0&0&0&1\\
		0&0&0&1&0
	\end{array} \right]}, \\
	CT=[0,     1,     0,     1,     1], \\
	FT=[0,     1,     1,     0,     0].\end{array}
\end{array}$${It is easy to check that the eigenvalues $\lambda=1$ and $\lambda=0$ are modal functionally observable, while $\lambda=-1$ is not. By Theorem \ref{theorem-func-observe}, $(A,C,F)$ is not functionally observable, but is functionally detectable.}
Furthermore, the observable canonical forms in the right hand side of (\ref{sys-decomp-observability}) corresponding to each Jordan block are respectively
\begin{equation}\label{blocks}\left[\begin{array}{c|c}
1&0\\
1&1\\
\hline
1&0\\
\hline
1&0
\end{array}\right],\left[\begin{array}{c|c}
-1&0\\
1&-1\\
\hline
1&0\\
\hline
0&1
\end{array}\right],\left[\begin{array}{c}
0\\
\hline
1\\
\hline
0
\end{array}\right].\end{equation}
Accordingly, the decomposition corresponding to (\ref{deeper-decomposition}) reads
$$\left[\begin{array}{ccc|cc}
	-1&0&0&0&0\\
	0&1&0&0&0\\
	0&0&0&0&0\\
	\hline
	1&0&0&-1&0 \\
	0&1&0& 0&1   \\
	\hline
	1&1 & 1& 0&0 \\
	\hline
	0&1 & 0& 1&0
	\end{array}\right],
$$which can be obtained by suitably permuting the blocks in (\ref{blocks}). It follows that $\tilde F_{\bar o1}\ne 0$ and $\tilde F_{\bar o2}=0$, which implies that $(A,C,F)$ is not functionally observable, but is functionally detectable, consistent with Theorem \ref{theorem-func-observe}.%
\end{example}}

Notably, the proposed {\emph{modal functional observability}} provides not only an eigenspace based characterization for functional observability, but also a complete criterion for functional detectability without using the observability decomposition as in Proposition \ref{decomposition-based-criterion}. Recall that the transformation matrix $T$ only depends on the eigenspace of $A$, independent of $C$ and $F$. This property facilitates greatly the design of measurement structure for a system to achieve functional observability. Say, by Theorem \ref{theorem-func-observe}, to
estimate a functional of state $z(t)=Fx(t)$, it suffices to construct $C_i$ with the same ${\rm row}(C_i)$ for each individual eigenvalue $\lambda_i$ of $A$ independently, such that (\ref{modal_functional}) holds. Then, the output matrix $C$ ensuring the functional observability of $(A,C,F)$ can be constructed as $C=[C_1,...,C_k]T^{-1}$ (to ensure that $C$ is real, each non-real eigenvalue and its complex conjugate need to be considered together; see section \ref{modal_sesnor_placement}). By accounting only unstable eigenvalues, a similar idea can be adopted to construct the measure structure to achieve functional detectability. In section \ref{modal_sesnor_placement}, we shall show this approach can lead to a closed-form solution to the minimum number of sensors needed to achieve functional observability for systems with diagonalizable state matrices (i.e. the algebraic multiplicity equals the geometric one for each eigenvalue of $A$; called {\emph{diagonalizable systems}} for short).



\begin{remark} \label{challenge-eigenspace}
	It is worth mentioning that \cite{jennings2011existence} initiated the idea of using the eigenspace of $A$ to characterize functional observability. They used the inner product of $C$ and $T_i$ to define what they referred to as `a matrix $C$ observing an eigenspace of $A$'. { However, except when $(J_i,C_i)$ is observable or $J_i$ is diagonal, ${\rm rank}[\lambda_iI-J_i;C_i]$ may fail to coincide with ${\rm rank}O(J_i,C_i)$. To see this, suppose $J_i$ consists of $r_i$ Jordan blocks, and the $j$th block has dimension $d'_j$, $j=1,...,r_i$. Let $C_{i}[j]$ be $j$th column of $C_i$, and define $l_{j}\doteq d'_1+\cdots+d'_{j-1}+1$ for $j=1,...,r_i$, where $l_0=0$.  $C_i[l_j]$ is called the $j$th leading vector of $C_i$.
		We then have ${\rm rank}[\lambda_iI-J_i;C_i]={\rm rank}[C_i[l_1],C_i[l_2],\cdots, C_i[l_{r_i}]]+(d_i-r_i)$. From it, ${\rm rank}[\lambda_iI-J_i;C_i]$ only depends on the leading vectors of $C_i$, while the information of non-leading vectors may be lost. However, the non-leading vectors do affect ${\rm rank}O(J_i,C_i)$ especially when $C_i[l_j]=0$ for $j=1,...,r_i$. Indeed, independent of $C_i$, we always have ${\rm rank}[\lambda_iI-J_i;C_i]\ge {\rm rank}(\lambda_iI-J_i)= d_i-r_i$. By contrast, depending on $C_i$, ${\rm rank}O(J_i,C_i)$ can range from $0$ to $d_i$.}

\end{remark}

By combining Lemma \ref{diagonal-modal} with Theorem \ref{theorem-func-observe},  we obtain the subsequent PBH-like condition for functional observability and detectability of diagonalizable systems. This condition, {previously asserted as necessary and sufficient for general systems as shown in \cite{moreno2001quasi,jennings2011existence,mohajerpoor2015minimal,Darouach2023FunctionalDA,montanari2022functional}},  is in fact valid for a subclass of systems. This  has also been  noticed by \cite{asymptotic2022Mohamed} recently. 


\begin{corollary}\label{functional-theorem-2}
	Suppose that $A$ is diagonalizable. The triple $(A,C,F)$ is functionally observable if and only if
	\begin{equation}\label{rank-criterion-observ} {\rm rank}  \left[\begin{array}{c}
	A-\lambda I_n \\
	C \\
	F
	\end{array}\right]={\rm rank}  \left[\begin{array}{c}
	A-\lambda I_n \\
	C
	\end{array}\right], \forall \lambda\in {\mathbb C}.\end{equation}	
	Moreover, the triple $(A,C,F)$ is functionally detectable if and only if
	\begin{equation}\label{rank-criterion-detect}
	{\rm rank}  \left[\begin{array}{c}
	A-\lambda I_n \\
	C \\
	F
	\end{array}\right]={\rm rank}  \left[\begin{array}{c}
	A-\lambda I_n \\
	C
	\end{array}\right], \forall \lambda\in {\mathbb C}_0^+.
	\end{equation}
\end{corollary}
{
	\begin{proof}	
	Notice that both sides of (\ref{rank-criterion-observ}) and (\ref{rank-criterion-detect}) lose full rank only when $\lambda$ is an eigenvalue of $A$. The required result then follows immediately from Theorem \ref{theorem-func-observe} and Lemma \ref{diagonal-modal}.
	\end{proof}	
}

{It is worth mentioning that when $A$ is not diagonalizable, conditions (\ref{rank-criterion-observ}) and (\ref{rank-criterion-detect}) are necessary for functional observability and functional detectability, respectively, which can be obtained from the proof for necessity parts of \cite[Theo. 4]{jennings2011existence} and \cite[Theo. 1]{Darouach2023FunctionalDA}.  However, their sufficiency cannot be guaranteed.} This can be seen from the following example.

\begin{example}\label{counter-example}
	Consider $(A,C,F)$ with\\ $A=\left[\begin{array}{ccc}
	0 & 1 & 0 \\
	0 & 0 & 1 \\
	0 & 0 & 0
	\end{array}\right]$, $C=\left[\begin{array}{ccc}
	0 & 0 & 1 \\
	\end{array}\right]$, $F=\left[\begin{array}{ccc}
	0 & 1 & 0 \\
	\end{array}\right].$
	Since ${\rm eig}(A)=\{0\}$, and
	\begin{equation} \label{fail-example} {\rm rank}  \left[\begin{array}{c}
	A \\
	C \\
	F
	\end{array}\right]=2={\rm rank}  \left[\begin{array}{c}
	A \\
	C
	\end{array}\right],\end{equation} conditions (\ref{rank-criterion-observ}) and (\ref{rank-criterion-detect}) are satisfied for this system. However, using Corollary \ref{fundamental-theorem}, it can be validated that ${\rm rank}[O(A,C);F]=2\ne {\rm rank}O(A,C)=1$,
	implying that $(A,C,F)$ is not functionally observable. Similarly, it follows from Theorem \ref{theorem-func-observe} that $(A,C,F)$ is not functionally detectable.
\end{example}

\section{Redefined structural functional observability} \label{Sec-IV}
In this section, we redefine SFO rigorously, and highlight its distinctions with structural observability.

Before presenting the definition of SFO, some {{notations and concepts}} are introduced.
Given a structured triple $(\bar A,\bar C, \bar F)$, with $\bar A\in \{0,*\}^{n\times n}$, $\bar C\in \{0,*\}^{p\times n}$, and $\bar F\in \{0,*\}^{r\times n}$, let $(A,C,F)$ be a realization of $(\bar A, \bar C, \bar F)$, i.e.
$A\in {\cal S}(\bar A)$, $C\in {\cal S}(\bar C)$, and $F\in {\cal S}(\bar F)$. An algebraic variety is defined as the set of solutions of a system of polynomial equations over real or complex numbers. A proper variety
of ${\mathbb R}^n$ is an algebraic variety that is not the whole space ${\mathbb R}^n$, which has Lebesgue measure zero in ${\mathbb R}^n$. Let $M$ be a matrix whose entries are polynomials of a set of free parameters $\theta$. The {\emph{generic rank}} of $M$, given by ${\rm grank}M$, is the maximum rank that $M$ can achieve as a function of $\theta$, which equals the rank $M$ can achieve for almost all values of $\theta$ \cite[page 38]{Murota_Book}. Note that the generic rank only concerns how $M$ is parameterized by $\theta$, but is independent of a specific value of $\theta$. 

\begin{definition}[Redefined SFO] \label{def-structural-functional} Let $\theta$ be a vector with dimension $d$ consisting of the parameters for the $*$ entries in the triple $(\bar A, \bar C, \bar F)$. $(\bar A, \bar C, \bar F)$ is said to be structurally functionally observable (SFO), if there is a proper variety ${\mathbb V}$ of ${\mathbb R}^d$, such that for all $\theta \in {\mathbb R}^d\backslash {\mathbb V}$ (in other words, for {\emph{almost all}} $\theta\in {\mathbb R}^d$), the corresponding realizations $(A, C, F)$ of $(\bar A, \bar C, \bar F)$ are functionally observable.
\end{definition}

SFO generalizes the concept structural observability in \cite{C.T.1974Structural}. The latter is defined for a structured pair $(\bar A,\bar C)$ as the property to allow an observable realization (see Section \ref{preliminary-structured-system}). It is known that observability is a generic property in the sense that either almost all realizations of $(\bar A, \bar C)$ are observable, or there is no observable realization. As shown in Corollary \ref{generic-property} of Section \ref{Sec-V}, functional observability is still a generic property but with subtly different behaviors: given a triple $(\bar A, \bar C,\bar F)$, either almost all its realizations are functionally observable, or almost all its realizations are functionally unobservable. In particular, unlike structural observability,  the existence of a functionally observable realization of $(\bar A, \bar C, \bar F)$ does {\emph{not}} imply that almost all its realizations are functionally observable.

The terminology SFO was first proposed in \cite{montanari2022functional}.
It is important to note that \cite{montanari2022functional} has defined it as the property of $(\bar A, \bar C, \bar F)$ to have a functionally observable realization. This definition cannot capture the embedded generic property and may lead to confusing results, as shown in the following example.

\begin{example} \label{illustrate-def-sfo}
	Consider $(\bar A, \bar C, \bar F)$ as
	$$\bar A=\left[\begin{array}{ccc}
	0 & 0 & 0\\
	* & 0 & 0\\
	0 & 0 & 0
	\end{array}\right],
	\bar C=\left[0,*,*\right],
	\bar F=\left[0,*,0 \right].$$
	There is a realization $(A,C,F)$ of $(\bar A, \bar C, \bar F)$ as
	$$ A=\left[\begin{array}{ccc}
	0 & 0 & 0\\
	1 & 0 & 0\\
	0 & 0 & 0
	\end{array}\right],
	C=\left[0,1,0\right],
	F=\left[0,1,0 \right]$$
	that is functionally observable (via Corollary \ref{fundamental-theorem}). However, assign parameters $(a_{21},c_{12},c_{13},f_{12})$ to the $*$ entries of $(\bar A, \bar C, \bar F)$ (the correspondence is indicated by the subscripts). We have
	$$\begin{array}{c}[O(A,C);F]=\left[\begin{array}{ccc}
	0 & c_{12} & c_{13}\\
	a_{21}c_{12} & 0 & 0\\
	0 & 0 & 0 \\
	\hline
	0 & f_{12} & 0
	\end{array}\right].\end{array}$$
	It is seen readily that for almost each $(a_{21},c_{12},c_{13},f_{12})\in {\mathbb R}^4$, ${\rm rank}O(A,C)=2<{\rm rank}[O(A,C);F]=3$, implying that $(A,C,F)$ is not functionally observable.
\end{example}

Example \ref{illustrate-def-sfo} also exhibits a dramatically different property between structural observability and SFO. That is, adding additional free  entries (parameter $c_{13}$ in this example) to $[\bar A, \bar C]$, corresponding to adding new connections/links among states or from states to outputs, may break the property of being SFO, which can be explained by the proposed criteria for SFO in the next section. By contrast, such a phenomenon never happens for structural observability.

\section{Criteria for SFO} \label{Sec-V}

Although a graph-theoretic necessary and sufficient condition for SFO was given in \cite{montanari2022functional}, this condition was based on condition (\ref{rank-criterion-observ}) that is applicable only to diagonalizable systems, thereby raising concerns about its validness for general systems. In this section, we propose complete graph-theoretic characterizations for SFO. Our results also have some interesting implication for structural target controllability.

\subsection{Linking and independent walk family}

Given a structured triple $(\bar A, \bar C, \bar F)$, the digraph ${\cal G}(\bar A,[\bar C;\bar F])$ is defined in a similar way to ${\cal G}(\bar A,\bar C)$ (see Section \ref{preliminary-structured-system}) by changing $\bar C$ to the {\emph{extended output matrix}} $[\bar C;\bar F]$, and accordingly changing the output vertices $Y=\{y_1,...,y_p\}$ to $\{y_1,...,y_p,y_{p+1},...,y_{p+r}\}$.  The {\emph{functional state set}} is defined as $X_F\doteq \{x_i: \bar F_{ji}\ne 0, \exists j
\in \{1,...,r\}\}$, corresponding to nonzero columns of $\bar F$, which is the set of state vertices that contribute directly to $z(t)$.

\begin{figure}
	\centering
	\includegraphics[width=3.3in]{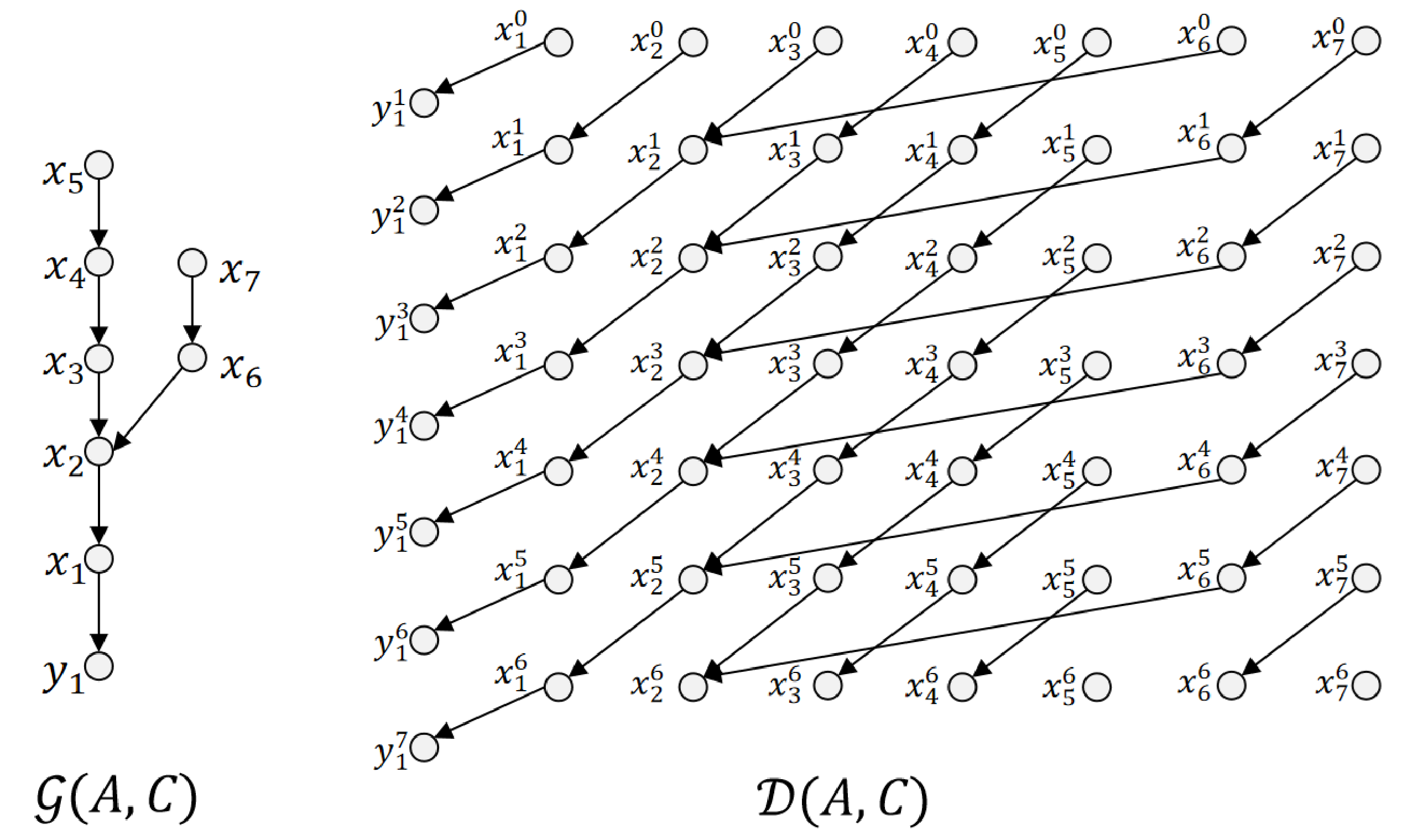}\\
	\caption{An example of ${\cal G}(\bar A, \bar C)$ and its associated ${\cal D}(\bar A, \bar C)$, where $(\bar A, \bar C)$ is given in Example \ref{counter-example-2}.} \label{SFO-example-digraph}
\end{figure}

For the structured matrix pair $(\bar A, \bar C)$, construct a digraph ${\cal D}(\bar A, \bar C)=(\bar X\cup \bar Y,E_{\bar X\bar X}\cup E_{\bar X\bar Y})$, where the vertex set $\bar X=\{x^{q}_{j}:j=1,...,n,q=0,...,n-1\}$, $\bar Y=\{y^{q}_{j}: j=1,...,p, q=1,...,n\}$, and the edge set $E_{\bar X\bar X}=\{(x^{q}_{j},x^{q+1}_{i}): \bar A_{ij}\ne 0, q=0,...,n-2\}$,
$E_{\bar X \bar Y}=\{(x_{j}^{q},y^{q+1}_{i}): \bar C_{ij}\ne 0, q=0,...,n-1\}$. {{An illustration of ${\cal G}(\bar A, \bar C)$ and ${\cal D}(\bar A, \bar C)$ can be found in Fig. \ref{SFO-example-digraph}.}} {As can be seen, $\bar X$ (resp. $\bar Y$) is obtained from $X$ (resp. $Y$) by duplicating each of its vertices $n$ times, and the $q$th copy of $x_j$ ($y_j$) is $x_j^q$ ($y_{j}^{q+1}$), $q=0,1,...,n-1$. Correspondingly,
	each edge $(x_i,x_j)\in E_{XX}$ corresponds to $n-1$ edges $(x_i^q,x_j^{q+1})\in E_{\bar X \bar X}$ for $q=0,\cdots, n-2$, and $(x_i,y_j)\in E_{XY}$ corresponds to $n$ edges $(x_i^q,y_j^{q+1})\in E_{\bar X \bar Y}$ for $q=0,\cdots,n-1$. Therefore, ${\mathcal D}(\bar A, \bar C)$ contains $n(n+p)$ vertices and $(n-1)|E_{XX}|+n|E_{XY}|$ edges. Here, the subscript $q$ indicates the order of a vertex in a path from $x_i$ to $y_j$ in ${\cal G}(\bar A, \bar C)$, such that any path from $x_i$ to $y_j$ with length $l\le n$ (with possibly repeated vertices, which will be called a walk) in ${\cal G}(\bar A, \bar C)$ corresponds to a unique path from $x^0_i$ to $y_j^l$ in ${\cal D}(\bar A, \bar C)$ (for example, the path $x_2\to x_1\to y_1$ in ${\cal G}(\bar A, \bar C)$ corresponds to $x_2^0\to x_1^1\to y_1^2$ in ${\cal D}(\bar A, \bar C)$ in Fig. \ref{SFO-example-digraph}).}  Such an acyclic digraph is called the {\emph{dynamic graph}} of $(\bar A, \bar C)$~\cite{poljak1990generic}.



 {Let $\hat Y\doteq \{y_j^{k}: j=1,...,p+r,k=1,...,n\}$.  The dynamic digraph ${\cal D}(\bar A,[\bar C;\bar F])$ on the vertex set $\bar X\cup \hat Y$ is defined in the same way as ${\cal D}(\bar A,\bar C)$ by considering $[\bar C;\bar F]$ as the output matrix corresponding to the extended output composed by the concatenation of $y$ and $z$.} Define another dynamic graph ${\cal D}(\bar A, \bar C, \bar F)\doteq (\bar X\cup \bar Y \cup \tilde Y,E_{\bar X\bar X}\cup E_{\bar X\bar Y}\cup E_{\bar F})$, with $\tilde Y=\{y^1_{p+1},...,y^1_{p+r}\}$, $E_{\bar F}=\{(x^0_i,y^1_{p+j}): \bar F_{ji}\ne 0\}$.  Let $\bar X_0=\{x_1^0,...,x_n^0\}$.  A linking in ${{\cal D}}(\bar A, \bar C)$ (resp., ${{\cal D}}(\bar A, [\bar C;
 \bar F])$ and  ${{\cal D}}(\bar A, \bar C, \bar F)$) is a collection of vertex-disjoint paths from $\bar X_0$ to $\bar Y$ (resp. $\hat Y$ and $\bar Y \cup \tilde Y$). The size of a linking  is the number of  paths it contains. A linking with the maximum size among all linkings of ${\cal D}(\bar A, \bar C)$ is called a {\emph{maximum linking}}.

The following lemma states the equivalence of several expressions for ${\rm grank}O(\bar A, \bar C)$. Notice that $n-{\rm grank}O(\bar A, \bar C)$ is exactly the generic dimension of unobservable subspaces of $(\bar A, \bar C)$.



\begin{lemma}\cite{poljak1990generic}\label{subspace-theorem}
	The following numbers are equal:
	\begin{itemize}
		\item[i)]  ${\rm grank}O(\bar A, \bar C)$.
		\item[ii)] The maximum size of a set $X_s\subseteq X$ that can be covered by a cactus configuration in ${\cal G}(\bar A, \bar C)$.
		\item[iii)] The maximum size of a linking in ${\cal D}(\bar A, \bar C)$.
	\end{itemize}
\end{lemma}

\begin{definition}[Independent walks]\label{def-walk} A walk $w=(v_0,v_1,...,v_k)$ in ${\cal G}(\bar A, \bar C)$ is a path (with possibly repeated vertices) from a state vertex $v_0\in X$ to an output vertex $v_k\in Y$, where $v_i$ is the $i$th vertex of the path, $i=0,...,k$, with $k\le n$ the length of $w$. Two walks $w=(v_0,v_1,...,v_k)$ and $w'=(v'_0,v'_1,...,v'_{k'})$ ($k,k'\le n$) are {\emph{independent}} if $v_i\ne v'_i$ for $i=0,1,...,\min \{k,k'\}$.  A collection of $k$ independent walks in which any two walks are independent is called an {\emph{independent walk family}} of size $k$. An independent walk family with the maximum size among all independent walk families of ${\cal G}(\bar A, \bar C)$ is called a maximum independent walk family.  
\end{definition}

\begin{definition} (Reached by an independent walk family) A vertex $x_i\in X$ is said to be reached by an independent walk family (in ${\cal G}(\bar A, \bar C)$), if it is the initial vertex of a walk in this independent walk family.
\end{definition}

{To illustrate the above definitions, consider ${\cal G}(\bar A, \bar C)$ in Fig. \ref{SFO-example-digraph}. An independent walk family with size $3$, for instance, could consist of three walks $(x_3,x_2,x_1,y_1)$, $(x_2,x_1,y_1)$, and $(x_1,y_1)$, which reaches vertices $\{x_3,x_2,x_1\}$.} By the constructions of ${\cal G}(\bar A, \bar C)$ and ${\cal D}(\bar A, \bar C)$, it follows that there is a one-to-one correspondence between an independent walk family of size $k$ in ${\cal G}(\bar A, \bar C)$ and a linking of size $k$ in ${\cal D}(\bar A, \bar C)$.  Indeed, a linking consisting of $2$ paths $(x^0_{i_0},x^1_{i_1},\cdots, x^{k_1-1}_{i_{k_1-1}},y^{k_1}_{i_{k_1}})$ and $(x^0_{i'_0},x^1_{i'_1},\cdots, x^{k_2-1}_{i'_{k_2-1}},y^{k_2}_{i'_{k_2}})$ corresponds to an independent walk family with $2$ independent walks $(x_{i_0},x_{i_1},\cdots, x_{i_{k_1-1}},y_{i_{k_1}})$ and $(x_{i'_0},x_{i'_1},\cdots, x_{i'_{k_2-1}},y_{i'_{k_2}})$, where $k_1,k_2\le n$. Hence, $x_i$ is reached by an independent walk family of size $k$ in ${\cal G}(\bar A, \bar C)$, if and only if $x_i^0$ is covered by a linking of size $k$ in ${\cal D}(\bar A, \bar C)$.

%


\subsection{Characterizations of structural functional observability}
\begin{proposition}\label{characterization-pro}
	The following statements are equivalent:
	\begin{itemize}
		\item[i)] The triple $(\bar A, \bar C, \bar F)$ is SFO.
		\item[ii)] ${\rm grank}[O(\bar A, \bar C);\bar F]={\rm grank}O(\bar A, \bar C)$.
		\item[iii)] ${\rm grank}[O(\bar A, \bar C); O(\bar A, \bar F)]={\rm grank}O(\bar A, \bar C)$.\footnote{Notice that ${\rm grank}[O(\bar A, \bar C); O(\bar A, \bar F)]\equiv {\rm grank}O(\bar A, [\bar C;\bar F])$. Hence, iii) is equivalent to ${\rm grank}O(\bar A, [\bar C;\bar F])={\rm grank}O(\bar A, \bar C)$. We adopt the expression of iii) because it is parallel to condition ii) of Corollary \ref{fundamental-theorem}.}
	\end{itemize}
\end{proposition}
\begin{proof} 
	ii)$\Rightarrow$i): Let the free parameters for the $*$ entries in $(\bar A,\bar C,\bar F)$ form a vector $\theta$ with dimension $d$. Suppose ii) holds.
	By the definition of generic rank, there is a proper variety ${\mathbb V}_1\subseteq {\mathbb R}^d$ such that for all $\theta \in {\mathbb R}^d\backslash {\mathbb V}_1$,
	${\rm rank} [O(A,C);F]={\rm grank} [O(\bar A,\bar C);\bar F ]$, where $(A,C,F)$ is obtained by substituting the value of $\theta$ into $(\bar A, \bar C, \bar F)$. Similarly, there is a proper variety ${\mathbb V}_2\subseteq {\mathbb R}^d$ such that for all $\theta \in {\mathbb R}^d\backslash {\mathbb V}_2$,
	${\rm rank} O(A,C)={\rm grank} O(\bar A,\bar C)$. Hence, for all $\theta \in {\mathbb R}^d\backslash ({\mathbb V}_2\cup {\mathbb V}_1)$,
	${\rm rank} [O(A,C);F]-{\rm rank} O(A,C)={\rm grank} [O(\bar A,\bar C);\bar F]-{\rm grank} O(\bar A,\bar C)=0$. This implies that, for all $\theta \in {\mathbb R}^d\backslash ({\mathbb V}_2\cup {\mathbb V}_1)$,
	the corresponding realizations $(A,C,F)$ are functionally observable. Since ${\mathbb V}_2\cup {\mathbb V}_1$ is still a proper variety of ${\mathbb R}^d$, by definition, $(\bar A, \bar C, \bar F)$ is SFO.
	
	i)$\Rightarrow$ii): If ii) does not hold, i.e., ${\rm grank}[O(\bar A, \bar C);\bar F]>{\rm grank}O(\bar A, \bar C)$, following an analysis as previous, it turns out that there is a property variety of ${\mathbb R}^d$, given by ${\mathbb V}_3$, such that for all $\theta \in {\mathbb R}^d\backslash {\mathbb V}_3$,
	${\rm rank} [O(A,C);F]={\rm grank} [O(\bar A,\bar C);\bar F]>{\rm grank} O(\bar A,\bar C)={\rm rank} [O( A, C)]$, i.e., the corresponding realizations $(A,C,F)$ are functionally unobservable. This indicates, there does not exist a proper variety ${\mathbb V}$ satisfying the condition in Definition \ref{def-structural-functional} (as the set of $\theta$ making the associated realization $(A,C,F)$ functionally observable must be a subset of ${\mathbb V}_3$). It follows that $(\bar A, \bar C, \bar F)$ cannot be SFO by Definition \ref{def-structural-functional}.
	
	Consequently, i) is equivalent to ii). {Notice that entries of $[O(\bar A, \bar C); O(\bar A, \bar F)]$ are polynomials of $\theta$.  By changing $[O(\bar A, \bar C); O(\bar A, \bar F)]$ to $O(\bar A,[\bar C;\bar F])$ and following a similar manner to i)$\Leftrightarrow$ii), we can prove i)$\Leftrightarrow$iii). Afterwards, ii)$\Leftrightarrow$iii) is immediate.}
\end{proof}

	From the proof of Proposition \ref{characterization-pro}, it is easy to get the following corollary on the generic property of functional observability.

\begin{corollary}\label{generic-property}
	Given a triple $(\bar A, \bar C,\bar F)$, either almost all its realizations are functionally observable (when $(\bar A, \bar C,\bar F)$ is SFO), or almost all its realizations are functionally unobservable (when $(\bar A, \bar C, \bar F)$ is not SFO).
\end{corollary}

By Proposition \ref{characterization-pro}, verifying SFO can reduce to checking whether condition iii) thereof is satisfied. From Lemma \ref{subspace-theorem}, both ${\rm grank}O(\bar A, [\bar C;\bar F])$ and ${\rm grank}O(\bar A, \bar C)$ can be attained by computing the maximum size of state vertices that can be covered by a cactus configuration or the maximum size of a linking in the respective graph. From \cite{poljak1990generic,murota1990note}, these two numbers can be computed in polynomial time by resorting to some graph-theoretic algorithms, more precisely, algorithms for the maximum weighted matching problem or the maximum flow problem in $O(n^3)$ time (see \citep[Theo. 6]{murota1990note}). {However, in the left-hand-side of condition iii), due to the fusion of $\bar C$ and $\bar F$ within a \textit{single} output matrix $[\bar C;\bar F]$ in ${\rm grank}[O(\bar A, \bar C); O(\bar A, \bar F)]$, the manner in which the inclusion of $\bar F$ impacts the maximum size of cactus configurations in ${\mathcal G}(\bar A, [\bar C;\bar F])$ remains implicit, obscuring the influence of the structure of $(\bar A, \bar C, \bar F)$ on the SFO.}

In what follows, we look deeper into SFO. In particular, the following Proposition \ref{single-pro} transforms the SFO of $(\bar A, \bar C, \bar F)$ into the SFO of $(\bar A, \bar C, \bar e_i)$, where $\bar e_i\in \{0,*\}^{1\times n}$ and its $i$th entry is $*$ and the rest zero, for each $x_i\in X_F$. 



\begin{proposition} \label{single-pro}
	The triple $(\bar A, \bar C, \bar F)$ is not SFO, if and only if there exists a $x_i\in X_{F}$ such that $(\bar A, \bar C, \bar e_i)$ is not SFO, with $\bar e_i$ defined above.
\end{proposition}

\begin{proof}
	The sufficiency is obvious, observing that when such $x_i$ exists, it holds that ${\rm grank}[O(\bar A, \bar C);\bar F]\ge {\rm grank}[O(\bar A, \bar C);\bar e_i]> {\rm grank} O(\bar A, \bar C)$.
	Now suppose the triple $(\bar A, \bar C, \bar F)$ is not SFO, i.e., ${\rm grank}[O(\bar A, \bar C);\bar F]> {\rm grank}O(\bar A, \bar C)$. Let $\bar f_i$ be the $j$th row of $\bar F$. By the property of generic rank function \cite{Murota_Book}, there must be a $j\in [r]$ such that ${\rm grank}[O(\bar A, \bar C);\bar f_j]> {\rm grank}\ O(\bar A, \bar C)$. Indeed, if this is not true, then one has ${\rm grank}[O(\bar A, \bar C);\bar F]= {\rm grank}\ O(\bar A, \bar C)$. Suppose $d_o={\rm grank}\ O(\bar A, \bar C)$. Then, it holds that ${\rm grank}[O(\bar A, \bar C);\bar f_j]=d_o+1$. This indicates there is a $(d_o+1)\times (d_o+1)$ sub-matrix in $[O(\bar A, \bar C);\bar f_j]$ whose determinant is not identically zero. In addition, this sub-matrix must contain a nonzero sub-row of $\bar f_j$. Since all nonzero entries in $\bar f_j$ take independent values, it follows that there exists a $x_i\in X_{\bar f_j}$ such that ${\rm grank}[O(\bar A, \bar C);\bar e_i]=d_o+1> {\rm grank}O(\bar A, \bar C)$. That is, $(\bar A, \bar C, \bar e_i)$ is not SFO. This proves the necessity.
\end{proof}

An interesting corollary of Proposition \ref{single-pro} is that SFO only depends on the set of states that contribute to $z(t)$ (i.e., the indices of nonzero columns in $\bar F$), but independent of how these states are linearly combined.

Let $d_o$, $\bar d_o$, and $\tilde d_o$ be the maximum size of a linking in ${\cal D}(\bar A, \bar C)$, ${\cal D}({\bar A, \bar C,\bar e_i})$, and ${\cal D}(\bar A,[\bar C;\bar e_i])$, respectively, $i\in \{1,...,n\}$, recalling ${\cal D}(\bar A, \bar C, \bar e_i)\doteq (\bar X\cup \bar Y\cup \{y^1_{p+1}\},E_{\bar X\bar X}\cup E_{\bar X\bar Y}\cup \{(x_i^0,y^1_{p+1})\})$.
By Lemma \ref{subspace-theorem}, $d_o={\rm grank}O(\bar A, \bar C)$, and equals the size of a maximum independent walk family of ${\cal G}(\bar A, \bar C)$.
Below, Proposition \ref{linking-simple-theo} establishes the relation between ${\rm grank}[O(\bar A, \bar C); \bar e_i]$ and the maximum size of a linking in ${\cal D}(\bar A, \bar C, \bar e_i)$. 

\begin{proposition} \label{linking-simple-theo}
	${\rm grank}[O(\bar A,\bar C);\bar e_i]=\bar d_o$.
\end{proposition}

\begin{proof}
	By the construction of dynamic graphs, ${\cal D}({\bar A, \bar C,\bar e_i})$ is obtained from ${\cal D}(\bar A, \bar C)$ by adding an edge $(x_i^0,y^1_{p+1})$ to the vertex $y^1_{p+1}$.
	We therefore have $d_o \le \bar d_o\le d_o+1$. By \citep[Lem 1 and Theo 2]{murota1990note}, we have ${\rm grank}[O(\bar A,\bar C);\bar e_i]\le \bar d_o$. Next, we consider two cases.
	If $\bar d_o=d_o$, since $d_o \le {\rm grank}[O(\bar A,\bar C);\bar e_i]\le \bar d_o$, we have ${\rm grank}[O(\bar A,\bar C);\bar e_i]=\bar d_o={\rm grank}O(\bar A,\bar C)$.
	If $\bar d_o=d_o+1$, since ${\cal D}({\bar A, \bar C,\bar e_i})$ is a subgraph of ${\cal D}(\bar A,[\bar C;\bar e_i])$, we have $\tilde d_o\ge \bar d_o= d_o+1$. {{Note that $\tilde d_o={\rm grank} O(\bar A,[\bar C;\bar e_i])$ by Lemma \ref{subspace-theorem}. It follows that ${\rm grank} O(\bar A,[\bar C;\bar e_i])\ge d_o+1>
	{\rm grank} O(\bar A,\bar C)$, which indicates that $(\bar A,\bar C,\bar e_i)$ is not SFO. By Proposition \ref{characterization-pro}, we have ${\rm grank} [O(\bar A,\bar C);\bar e_i]>
	{\rm grank} O(\bar A,\bar C)=d_o$.}} This restricts that ${\rm grank} [O(\bar A,\bar C);\bar e_i]=d_o+1=\bar d_o$, as ${\rm grank} [O(\bar A,\bar C);\bar e_i]\le {\rm grank} O(\bar A,\bar C)+1$. Therefore, in both cases, ${\rm grank}[O(\bar A,\bar C);\bar e_i]=\bar d_o$.
\end{proof}

\begin{remark}
	It is important to note that, in general, a linking terminated at $\bar Y'\subseteq \bar Y$ in ${\cal D}(\bar A, \bar C)$ does not imply that the columns of $O(\bar A,\bar C)$ corresponding to $\bar Y'$ have full column rank generically, even when each sensor measures only one state. In addition, the maximum size of a linking in ${\cal D}(\bar A,\bar C, \bar F)$ can only upper bound ${\rm grank}[O(\bar A, \bar C);\bar F]$, with no guarantee to coincide with ${\rm grank}[O(\bar A, \bar C);\bar F]$ (see \cite{murota1990note}). It is still open whether ${\rm grank}[O(\bar A, \bar C);\bar F]$ can be computed in polynomial time \cite{murota1990note,czeizler2018structural}. The key to establishing the equality in Proposition \ref{linking-simple-theo} when $\bar F=\bar e_i$ is the utilization of Proposition \ref{characterization-pro}, i.e., the equivalence between conditions ii) and iii) thereof.
\end{remark}



Based on Propositions \ref{characterization-pro}, \ref{single-pro}, and \ref{linking-simple-theo}, we give a complete graph-theoretic characterization of SFO as follows.

\begin{theorem}[Graph-theoretic characterization of SFO] \label{linking-theorem}
	The triple $(\bar A,\bar C,\bar F)$ is SFO, if and only if the functional state set $X_F$ does not contain any vertex that is not reached by a
	maximum independent walk family in ${\cal G}(\bar A, \bar C)$. In other words, $(\bar A,\bar C,\bar F)$ is SFO, if and only if each state vertex $x_i\in X_F$ is {\emph{reached by every}} maximum independent walk family (of size $d_o$) in ${\cal G}(\bar A, \bar C)$, or put differently, $x_i^0$ for each $x_i\in X_F$ is covered by {\emph{every}} maximum linking of ${\cal D}(\bar A, \bar C)$.
\end{theorem}

\begin{proof}
	Necessity: Suppose there is a $x_i\in X_F$ such that $x_i^0$ is not covered by a maximum linking $L$ of ${\cal D}(\bar A,\bar C)$. Then, by the construction of ${\cal D}(\bar A,\bar C,\bar e_i)$, the size of a maximum linking of ${\cal D}(\bar A,\bar C,\bar e_i)$ is $d_o+1$ (indeed, $L\cup \{(x_i^0,y^1_{p+1})\}$ forms a maximum linking of size $d_o+1$). From Proposition \ref{linking-simple-theo},
	${\rm grank}[O(\bar A,\bar C);\bar e_i]=d_o+1>{\rm grank}O(\bar A,\bar C)$, indicting that $(\bar A, \bar C, \bar F)$ is not SFO according to Proposition \ref{single-pro}.
	
	Sufficiency: For any $x_i\in X_F$, since $x_i^0\in X_F$ is covered by every maximum linking of ${\cal D}(\bar A, \bar C)$, it follows that the maximum size of a linking in ${\cal D}(\bar A,\bar C,\bar e_i)$ is still $d_o$.
	Indeed, if there is a linking of size $d_o+1$ in ${\cal D}(\bar A,\bar C,\bar e_i)$, given by $L'$, then the path $(x_i^0,y^1_{p+1})$ must be contained in $L'$ (otherwise the maximum size of a linking in ${\cal D}(\bar A,\bar C)$ is $d_o+1$). As a result, the linking $L$ obtained from $L'$ after removing the path $(x_i^0,y^1_{p+1})$ is of size $d_o$, which does not cover $x_i^0$, causing a contradiction. By Proposition \ref{linking-simple-theo},  ${\rm grank}[O(\bar A,\bar C);\bar e_i]=d_o={\rm grank}O(\bar A,\bar C)$, indicting that $(\bar A, \bar C, \bar e_i)$ is SFO. Since this is true for each $x_i\in X_F$,  $(\bar A, \bar C, \bar F)$ is SFO by Proposition \ref{single-pro}. 
\end{proof}

{Notice that when $X_F=X$, each $x_i\in X$ is reached by every maximum independent walk family in ${\cal G}(\bar A, \bar C)$, if and only if any maximum independent walk family in ${\cal G}(\bar A, \bar C)$ reaches all state vertices, i.e., $d_o=n$, which reduces to exactly the condition in Lemma \ref{structural-observability-theorem} for structural observability. } 

\begin{remark}It is worth noting that a set $X_s\subseteq X$ of size $d_o$ covered by a cactus configuration must be reached by a maximum independent walk family, while the converse is not necessarily true (see Example \ref{counter-example-2}). Hence, even though a vertex $x_i\in X$ is covered by every cactus configuration that covers $d_o$ state vertices, it does not imply that $(\bar A,\bar C, \bar e_i)$ is SFO. \end{remark}

\begin{remark}
	To verify whether a given $x_i\in X_F$ is reached by \emph{every} maximum independent walk family of ${\cal G}(\bar A, \bar C)$, we can determine the maximum size of a linking from $\bar X_0\backslash \{x_i^0\}$ to $\bar Y$ in ${\cal D}(\bar A, \bar C)$. If the size is $d_o$, meaning a linking of size $d_o$ not covering $x_i^0$ exists, the answer is NO; otherwise, YES.
\end{remark}

\begin{remark}
	From Theorem \ref{linking-theorem}, adding a link $e\in X\times (X \cup Y)$ to ${\cal G}(\bar A, \bar C)$ may either increase $d_o$ or generate new maximum independent walk families, making a vertex $x_i\in X_F$   that is reached by every maximum independent walk family in ${\cal G}(\bar A, \bar C)$ not necessarily satisfy the same property in  the new system digraph. This is consistent with the fact that adding new links to ${\cal G}(\bar A, \bar C)$ may destroy SFO (see Example \ref{illustrate-def-sfo}).
\end{remark}

Under the additional assumptions that each row of $\bar F$ has only one $*$ entry and ${\rm grank}[\bar C;\bar F]=p+r$,	it is claimed in \cite{montanari2022functional} that $(\bar A, \bar C, \bar F)$ is SFO, if and only if a) every state vertex $x_i\in X_F$ is output-reachable, and b) no $x_i\in X_F$ is contained in a minimal dilation of ${\cal G}(\bar A,
\bar C)$ ({\emph{labeled conditions a) and b)}). Here, a subset $X'\subseteq X$ is a dilation of ${\cal G}(\bar A, \bar C)$ if $|T(X')|<|X|$, where $T(X')=\{v: (x,v)\in E_{XX}\cup E_{XY}, x\in X'\}$ is the set of out-neighbors of vertices in $X'$, and a dilation $X'$ is minimal if any proper subset $X''\subsetneqq X'$ is not a dilation. The following example shows that these conditions are not sufficient for SFO. 
	
	
	\begin{example} \label{counter-example-2}
		Consider a pair $(\bar A, \bar C)$ as follows
			{$$\bar A=\left[\begin{array}{ccccccc}
			0 & * & 0 & 0 & 0 & 0 & 0\\
			0 & 0 & * & 0 & 0 & * & 0\\
			0 & 0 & 0 & * & 0 & 0 & 0\\
			0 & 0 & 0 & 0 & * & 0 & 0\\
			0 & 0 & 0 & 0 & 0 & 0 & 0\\
			0 & 0 & 0 & 0 & 0 & 0 & *\\
			0 & 0 & 0 & 0 & 0 & 0 & 0\\
			\end{array}\right], \bar C=[*,0,0,0,0,0,0],$$}whose associated ${\cal G}(\bar A, \bar C)$ is given in Fig. \ref{SFO-example-digraph}. It is easy to see that every state vertex is output-reachable, and there is only one minimal dilation $\{x_3,x_6\}$. Hence, the criterion in \cite{montanari2022functional} gives that $(\bar A, \bar C, \bar e_i)$ is SFO for each $i\in \{1,2,4,5,7\}$. However, using Proposition \ref{characterization-pro} directly yields that $(\bar A, \bar C, \bar e_i)$ is not SFO for $i\in \{4,7\}$.
		Now let us use Theorem \ref{linking-theorem} to reproduce this result. Notice that the maximum size of a state set that is covered by a cactus configuration is $5$, and there is only one such set, being $\{x_1,x_2,...,x_5\}$ (covered by an output-stem $(x_5,x_4,...,x_1,y_1)$). This set corresponds to a maximum independent walk family of ${\cal G}(\bar A, \bar C)$, consisting of $5$ independent paths $(x_1,y_1)$, $(x_2,x_1,y_1)$, $(x_3,x_2,x_1,y_1)$, $(x_4,x_3,x_2,x_1,y_1)$, and $(x_5,x_4,x_3,x_2,x_1,y_1)$, which does not reach $\{x_6,x_7\}$. Hence, Theorem \ref{linking-theorem} yields that $(\bar A, \bar C, \bar e_i)$ is not SFO for $i\in \{6,7\}$. Moreover, there is another maximum independent walk family consisting of $5$ paths $(x_1,y_1)$, $(x_2,x_1,y_1)$, $(x_6,x_2,x_1,y_1)$, $(x_7,x_6,x_2,x_1,y_1)$, and $(x_5,x_4,x_3,x_2,x_1,y_1)$, which does not reach $\{x_3,x_4\}$. As a result, $(\bar A, \bar C, \bar e_i)$ is not SFO for $i\in \{3,4\}$ from Theorem \ref{linking-theorem}. Furthermore, it can be verified that every maximum independent walk family of ${\cal G}(\bar A, \bar C)$ reaches $\{x_1,x_2,x_5\}$. Therefore, $(\bar A, \bar C, \bar e_i)$ is SFO for $i\in \{1,2,5\}$, which can also be validated via Proposition \ref{characterization-pro}.
	\end{example}

The following corollary, derived from Theorem \ref{linking-theorem}, shows that the above-mentioned two conditions, namely, conditions a) and b), are necessary for SFO. Note that this necessity does not require any additional constraint on the structure of $\bar C$ and $\bar F$.

\begin{corollary}\label{corollary-2-condition}
	If $(\bar A, \bar C, \bar F)$ is SFO, then every state vertex $x_i\in X_F$ is output-reachable, and no $x_i\in X_F$ is contained in a minimal dilation of ${\cal G}(\bar A,
	\bar C)$.
\end{corollary}

\begin{proof}
	If there is a state $x_i\in X_F$ that is not output-reachable, $x_i$ cannot be reached by any independent walk family of ${\cal G}(\bar A, \bar C)$. By Theorem \ref{linking-theorem}, $(\bar A, \bar C, \bar F)$ is not SFO. Now consider the second condition.  Assume that $(\bar A, \bar C, \bar F)$ is SFO. Suppose $x_i\in X_F$ is contained in some minimal dilation $X'$. Since $X'$ is a minimal dilation, by definition, for any proper subset $X''\subsetneqq X'$, $|T(X'')|\ge |X''|$ and $|T(X')|<|X'|$. It turns out that for any $x\in X'$, $|T(X'\backslash \{x\})|=|X'|-1$ and therefore $T(X'\backslash \{x\})=T(X')$.  Consider
		any $X_s\subseteq T(X')$ and any $x\in X'$. Define $\bar T(X_s)=\{\bar x\in X'\backslash \{x\}:(\bar x,x')\in E_{XX}\cup E_{XY},x'\in X_s\}$ as the set of in-neighbors of $X_s$ that belong to $X'\backslash \{x\}$. We are to show that $|\bar T(X_s)|\ge |X_s|$. Indeed, when $T(\{x\})\cap X_s\ne \emptyset$, if $|\bar T(X_s)|<|X_s|$, then $|T(X'\backslash (\{x\}\cup \bar T(X_s)))|=|T(X')\backslash X_s|=|T(X')|-|X_s|=|T(X'\backslash \{x\})|-|X_s|<|X'\backslash \{x\}|-|\bar T(X_s)|=|X'\backslash (\{x\}\cup \bar T(X_s))|$, implying that $X'\backslash (\{x\}\cup \bar T(X_s))$ is a dilation, contradicting the fact that $X'$ is a minimal dilation. When $T(\{x\})\cap X_s= \emptyset$, if $|\bar T(X_s)|<|X_s|$, then $|T(X'\backslash \bar T(X_s))|=|T(X')\backslash X_s|=|T(X')|-|X_s|<|X'|-|\bar T(X_s)|=|X'\backslash \bar T(X_s)|$, meaning that $ X'\backslash \bar T(X_s)$ is dilation, again causing a contradiction.

 Let $\bar X_0'=\{x_j^0: x_j\in X'\}$ and $\bar X_1'=\{x_j^1: x_j\in T(X')\}$.  Let $L$ be a maximum linking of ${\cal D}(\bar A, \bar C)$ that covers $x_i^0$, and $\bar X_L$ be the set of vertices in $L$.  Define $\bar X_0''=\bar X_L\cap \bar X_0'$ and $\bar X_1''=\bar X_L\cap \bar X_1'$. Then, $|\bar X_0''|=|\bar X_1''|$. It follows from the above analysis and the correspondence between ${\cal G}(\bar A, \bar C)$ and ${\cal D}(\bar A, \bar C)$ that there are $|\bar X_1''|$ disjoint edges between $\bar X_0'\backslash \{x_i^0\}$ and $\bar X_1''$. It is not difficult to see that, by changing the edges between $\bar X_0''$ and $\bar X_1''$ in $L$ to the edges between $\bar X_0'\backslash \{x_i^0\}$ and $\bar X_1''$, a new maximum linking $L'$ can be obtained from $L$ that does not cover $x_i^0$. This contradicts the fact that every maximum linking of ${\cal D}(\bar A, \bar C)$ covers $x_i^0$ from Theorem \ref{linking-theorem}. Hence,  $(\bar A, \bar C, \bar F)$ cannot be SFO. 
\end{proof}

{
\begin{remark} Recently, \cite{zhang2024diagonalizability} introduced the concept {\emph{structurally diagonalizable systems}}, referring to structured systems of which almost all realizations are diagonalizable (for detailed characterizations, see \cite[Sec. 4]{zhang2024diagonalizability}). They demonstrated that conditions a) and b) obtained in \cite{montanari2022functional} are actually necessary and sufficient for the SFO of this class of systems.
\end{remark}}


%


\begin{corollary} \label{self-loop-coro}
	If every state $x_i\in X$ has a self-loop, then $(\bar A, \bar C, \bar F)$ is SFO, if and only if each state $x_i\in X_F$ is output-reachable in ${\cal G}(\bar A, \bar C)$.
\end{corollary}
\begin{proof}
	The necessity of output-reachability comes from Corollary \ref{corollary-2-condition}. For sufficiency, let $X_{\rm re}\subseteq X$ be the set of output-reachable state vertices. Denote ${\cal G}'(\bar A, \bar C)$ the subgraph of ${\cal G}(\bar A, \bar C)$ induced by $X_{\rm re}\cup Y$. It turns out that ${\rm grank}O(\bar A, \bar C)$ equals $|X_{\rm re}|$ from Lemma \ref{subspace-theorem}. Hence, each $x_i\in X_{\rm re}$ is reached by every maximum independent walk family with size $|X_{\rm re}|$ of ${\cal G}'(\bar A, \bar C)$. Since $X_F\subseteq X_{\rm re}$, $(\bar A, \bar C, \bar F)$ is SFO from Theorem \ref{linking-theorem}.
\end{proof}

\subsection{Implication to structural target controllability}
A triple $(A,B,F)$ is said to be output controllable, if for any pair $z_0,z_f\in {\mathbb R}^r$, there exists a finite time $T$ and an input $u(t)$ over $[0,T]$ that steers $z(t)=Fx(t)$ from the initial vector $z_0=Fx(0)$ to the final vector $z_f=Fx(T)$ \cite{Modern_Control_Ogata}. Let $Q(A,B)=[B,AB,...,A^{n-1}B]$ be the controllability matrix of $(A,B)$. $(A,B,F)$ is output controllable, if and only if $FQ(A,B)$ has full row rank \cite{murota1990note}. A structured triple $(\bar A,\bar B, \bar F)$ is said to be structurally output controllable, if there exists a realization of it that is output controllable \cite{murota1990note}. Particularly, when $\bar F=\bar I_{S}$, $S\subseteq [n]$ ($\bar I$ is an $n\times n$ diagonal matrix with diagonal entries being $*$), if $(\bar A,\bar B, \bar F)$ is structurally output controllable, $(\bar A, \bar B, S)$ is called structurally target controllable \cite{gao2014target}.  Unlike the duality between controllability and observability, criteria for (structural) output/target controllability and SFO are not simply dual to each other; {that is, the (structural) output controllability of $(A,B,F)$ is not equivalent to the (structural) functional observability of $(A^{\intercal}, B^{\intercal}, F)$}.\footnote{{We remark that some weak and strong duality between output controllability and functional observability (i.e., one implies the other under certain conditions) are recently established in \cite{montanari2024duality}.}} Specially, our results show that SFO can be verified in polynomial time, while the same verification problem for structural output/target controllability is a long-standing open problem \cite{murota1990note,czeizler2018structural}. Interestingly, our results  reveal that a special case of the latter problem is polynomially solvable.

\begin{corollary}[Polynomially verifiable structural target controllability] \label{coro-target-controllable}
	There is a polynomial-time algorithm that can verify whether a triple $(\bar A,\bar B, S)$ is structurally target controllable for any $S\subseteq [n]$ with $|S|=n-1$.
\end{corollary}
\begin{proof} {Notice that $(\bar A,\bar B, S)$ is structurally target controllable, if and only if $\bar I_{S}Q(\bar A,\bar B)=Q(\bar A,\bar B)_S$ has full row generic rank \cite{gao2014target}. Therefore,} it suffices to prove that ${\rm grank}Q(\bar A,\bar B)_S$ can be determined in polynomial time. Let $i=[n]\backslash S$. Notice that ${\rm grank}[Q(\bar A,\bar B),\bar e_i^{\intercal}]=1+{\rm grank}Q(\bar A,\bar B)_S$.
By transposing all matrices in Proposition \ref{linking-simple-theo},  we can determine  ${\rm grank}[Q(\bar A,\bar B),\bar e_i^{\intercal}]$ in polynomial time by computing the maximum size of a linking in the dynamic graph associated with $[Q(\bar A,\bar B),\bar e_i^{\intercal}]$. As a result, ${\rm grank}Q(\bar A,\bar B)_S$ can be determined in polynomial time.
\end{proof}

\begin{remark} \label{hard}
When $S\subseteq [n]$ is such that $|S|\le n-2$, upon defining $\bar S=[n]\backslash S$,  we can determine whether ${\rm grank}[Q(\bar A,\bar B),\bar I_{\bar S}^{\intercal}]>{\rm grank}Q(\bar A,\bar B)$ by verifying the SFO of $(\bar A^{\intercal}, \bar B^{\intercal}, I_{\bar S})$ in polynomial time. In particular, if $(\bar A^{\intercal}, \bar B^{\intercal}, I_{\bar S})$ is not SFO, then ${\rm grank}[Q(\bar A,\bar B),\bar I_{\bar S}^{\intercal}]>{\rm grank}Q(\bar A,\bar B)$. It follows that whether ${\rm grank}[Q(\bar A,\bar B),\bar I_{\bar S}^{\intercal}]$ is within the range $[{\rm grank}Q(\bar A,\bar B)+1,{\rm grank}Q(\bar A,\bar B)+|\bar S|]$ can be determined in polynomial time. With ${\rm grank}Q(\bar A,\bar B)_S={\rm grank}[Q(\bar A,\bar B),\bar I_{\bar S}^{\intercal}]-|\bar S|$, we obtain that determining whether ${\rm grank}Q(\bar A,\bar B)_S$ is within the range $[{\rm grank}Q(\bar A,\bar B)+1-|\bar S|, {\rm grank}Q(\bar A,\bar B)]$ is achievable in polynomial time. {Additionally, observe that ${\rm grank}[O(\bar A,\bar C);\bar I_{\bar S}]={\rm grank}[Q(\bar A^\intercal,\bar C^\intercal),\bar I_{\bar S}^{\intercal}]={\rm grank}Q(\bar A^\intercal,\bar C^\intercal)_S+|\bar S|$.  The previous results hold for computing ${\rm grank}[O(\bar A,\bar C);\bar I_{\bar S}]$ when verifying SFO via condition ii) of Proposition \ref{characterization-pro}.}
\end{remark}

\section{Minimal sensor placement for functional observability}\label{Sec-VI}
In this section, we consider the problems of selecting the minimum number of sensors from a prior set of possible placements to achieve functional observability and SFO.
These problems may be desirable when one aims to estimate linear functions of states while minimizing sensor usage; {see \cite{pham2015load} for an application of these problems in the load frequency control of power systems using distributed functional observers.}  	
Based on the previous results, we will prove their NP-hardness and propose efficient algorithms to find approximation solutions. We also identify a special case that allows a closed-form solution, which generalizes the maximum geometric multiplicity result for the unconstrained minimal observability problem  \cite{Tong2017Minimal}.   

Let $C\in {\mathbb R}^{p\times n}$ be the output matrix corresponding to the available sensor set. 	Formally, given a triple $(A,C,F)$ ($(\bar A, \bar C, \bar F)$), the minimal sensor placement problems for functional observability and SFO can be respectively formulated as
\begin{equation}\begin{array}{l}
\min \limits_{S\subseteq [p]} |S| \tag{${\cal P}_1$}\label{A1} \\
{\rm s.t.} \ (A,C_{S},F) \ {\rm functionally \ observable}
\end{array}\end{equation}and
\begin{equation}\begin{array}{l}
\min \limits_{S\subseteq [p]} |S| \tag{${\cal P}_2$}\label{A2} \\
{\rm s.t.} \ (\bar A,\bar C_{S},\bar F) \ {\rm structurally \ functionally \ observable}.
\end{array}\end{equation}

\subsection{Computational complexity}
\begin{proposition}
	Both problems ${\cal P}_1$ and ${\cal P}_2$ are NP-hard, even with dedicated sensors, i.e., there is only one nonzero entry (resp. $*$ entry) in each row of $C$ (resp. $\bar C$).
\end{proposition}
\begin{proof}
	By setting $F=I_n$, ${\cal P}_1$ collapses to the minimal observability problem considered in \cite{A.Ol2014Minimal}, which is shown to be NP-hard even when $C=I_n$. Therefore, ${\cal P}_1$ is NP-hard even with dedicated sensors.
	
	To prove the NP-hardness of ${\cal P}_2$, we reduce the set cover problem to an instance of ${\cal P}_2$. {See Fig. \ref{np-hard-illustration} for illustration of such a reduction.} Given a ground set $S=\{1,2,...,r\}$ and a collection of its subsets $S_i\subseteq S$, $i=1,...,p$, the set cover problem seeks to determine a subset ${\cal I}\subseteq [p]$ with the minimum $|{\cal I}|$ such that $\bigcup\nolimits_{i\in {\cal I}}S_i=S$. This problem is NP-hard \cite{garey1979computers}. Consider the aforementioned set cover problem with ground set $S$ and its $p$ subsets $S_i\subseteq S$. Without loss of generality, assume that $\bigcup\nolimits_{i\in [p]}S_i=S$. Construct ${\cal G}(\bar A, \bar C)$ with $X=\{x_1,...,x_{r+p}\}$, $Y=\{y_1,...,y_{r+p}\}$, $E_{XX}=\{(x_j,x_{r+i}): j\in S_i, i=1,...,p,j=1,...,r)\}\cup \{(x_i,x_i): i=1,...,r+p\}$, and $E_{XY}=\{(x_i,y_i): i=1,...,r+p\}$. In other words, each state vertex of $(\bar A, \bar C)$ has a self-loop, there is an edge $(x_j,x_{r+i})$ if $j\in S_i$, and a dedicated sensor is available for each state. Let the functional state set $X_F=\{x_1,...,x_r\}$ (i.e., $\bar F={\bf col}\{\bar e_1,...,\bar e_r\}$).
	
	It is claimed that, there is a ${\cal I}\subseteq [p]$ with $|{\cal I}|\le k$ such that $\bigcup\nolimits_{i\in {\cal I}}S_i=S$, if and only if ${\cal P}_2$ on $(\bar A, \bar C, \bar F)$ has a solution ${\cal J}\subseteq [r+p]$ with $|{\cal J}|\le k$. For the one direction, if a set $|{\cal I}|\le k$ makes $\bigcup\nolimits_{i\in {\cal I}}S_i=S$, then ${\cal J}=\{r+i: i\in {\cal I}\}$ makes every $x_i\in X_F$ output reachable in ${\cal G}(\bar A, \bar C_{\cal J})$. By Corollary \ref{self-loop-coro}, $(\bar A, \bar C_{\cal J}, \bar F)$ is SFO with $|{\cal J}|\le k$. For the other direction, suppose ${\cal J}\subseteq [r+p]$ with $|{\cal J}|\le k$ makes $(\bar A, \bar C_{\cal J}, \bar F)$ SFO. From Corollary \ref{self-loop-coro}, this requires
	every $x_i\in X_F$ to be output-reachable in ${\cal G}(\bar A, \bar C_{\cal J})$, equivalently, $\bigcup\nolimits_{i\in {\cal J}, i>r}S_i\cup \{i: i\in {\cal J}, i\le r\}=S$. For any $i\in {\cal J}$ with $i\le r$, let $l(i)\in [p]$ be such that $i\in S_{l(i)}$. Define ${\cal I}=\{i-r: i\in {\cal J}, i>r\}\cup \{l(i): i\in {\cal J}, i\le r\}$. It follows that $\bigcup\nolimits_{i\in {\cal I}}S_i=S$, i.e., there is a solution to the set cover problem with $|{\cal I}|\le |{\cal J}|\le k$. Since deciding whether there is a solution ${\cal I}$ with $|{\cal I}|\le k$ for the set cover problem is NP-complete, we conclude that ${\cal P}_2$ is NP-hard.
\end{proof}

	\begin{figure}
	\centering
	\includegraphics[width=2.6in]{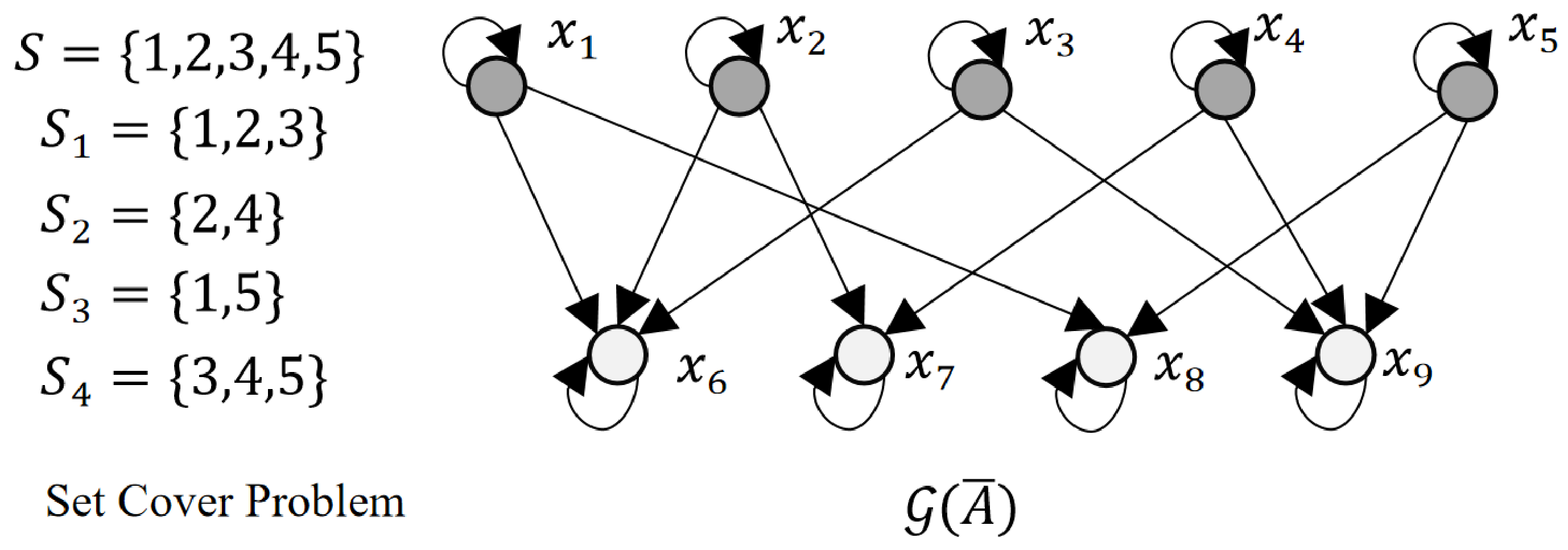}\\
	\caption{Illustration of the construction from a set cover problem to an instance of ${\cal P}_2$. Filled nodes represent functional states. Since
each state has a dedicated sensor, only ${\cal G}(\bar A)$ is present. } \label{np-hard-illustration}
\end{figure}


A problem akin to $\mathcal{P}_1$, labeled as $\mathcal{P}_3$, involves identifying the smallest subset of sensors $S \subseteq [p]$ that renders the resulting $(A, C_S, F)$ functionally detectable. Notably, when $F$ is the identity matrix $I_n$ and $A$ lacks stable eigenvalues, this problem reduces to the minimal observability problem within the class of systems devoid of stable eigenvalues, which is shown to be NP-hard in \cite{A.Ol2014Minimal}. Hence, ${\cal P}_3$ is also NP-hard.
{
\begin{remark}
When $\bar F= \bar I_n$,  ${\cal P}_2$ reduces to the problem of determining the minimum number of dedicated sensors to achieve {\emph{structural observability}}. Remarkably, this problem admits a polynomial-time solution \cite{Ramos2022AnOO} (this does not contradict the NP-hardness of ${\cal P}_1$ when $F=I_n$, since structural observability concerns generic system realizations rather than {\emph{every}} realization as in ${\cal P}_1$). The NP-hardness of ${\cal P}_2$ with dedicated sensors sharply contrasts with this fact.
\end{remark}}

\subsection{Supermodularity and greedy heuristics}
Despite the NP-hardness, in the following, by establishing supermodular functions that map the possible selections to some scalar functions, we show problems ${\cal P}_1$ and ${\cal P}_2$ can be approximated using greedy heuristics with provable guarantees. Please refer to Section \ref{preliminary-submodular} for definitions of supermodularity and monotonicity.

\begin{proposition} \label{sup-numerical}
	Define set functions $f(S)\doteq {\rm rank} [O(A,C_S);F]-{\rm rank} O(A,C_S)$ and $\bar f(S)\doteq {\rm rank} [O(A,[C_S;F])]-{\rm rank} O(A,C_S)$, $S\subseteq [p]$. Both $f(S)$ and $\bar f(S)$ are supermodular and monotone decreasing.
\end{proposition}

\begin{proof}
	Note ${\rm rank} [O(A,C_S);F]={\rm rank} O(A,C_S)+ {\rm rank} F- {\rm dim}({\cal R}(O(A,C_S))\cap {\cal R}(F))$. This yields
	$$f(S)={\rm rank} F- {\rm dim}({\cal R}(O(A,C_S))\cap {\cal R}(F)).$$
	Since ${\rm rank} F$ is constant and ${\rm dim}({\cal R}(O(A,C_S))\cap {\cal R}(F))$ increases with $S$, $f(S)$ is monotone decreasing. Moreover, to show the supermodularity of $f(S)$, it suffices to show the submodularity of ${\rm dim}({\cal R}(O(A,C_S))\cap {\cal R}(F))$. For any $a\in [p]$, define the gain function $h_a(S)={\rm dim}({\cal R}(O(A,C_{S\cup \{a\}}))\cap {\cal R}(F))-{\rm dim}({\cal R}(O(A,C_S))\cap {\cal R}(F))$.
	Observing that
	\begin{align*}
	&{\rm dim}({\cal R}(O(A,C_{S\cup \{a\}}))\cap {\cal R}(F))\\
	=&{\rm dim}({\cal R} (O(A,C_{S}))\cap {\cal R} (F))\\&+{\rm dim} ({\cal R} (O(A,C_{\{a\}}))\cap {\cal R} (F))\\ &-{\rm dim} ({\cal R} (O(A,C_{S}))\cap {\cal R} (F) \cap O(A,C_{\{a\}})),
	\end{align*}
	we have
	\begin{align*} h_a(S)&={\rm dim} ({\cal R} (O(A,C_{\{a\}}))\cap {\cal R} (F))\\ &- {\rm dim} ({\cal R} (O(A,C_{S}))\cap {\cal R} (F) \cap O(A,C_{\{a\}})).\end{align*}
	Notice that the first term in the above $h_a(S)$ is constant and the second term increases with $S$ since ${\rm dim} (O(A,C_{S}))$ does. Hence, $h_a(S)$ deceases with $S$.
	This proves the submodularity of ${\rm dim}({\cal R}(O(A,C_S))\cap {\cal R}(F))$ on $S\subseteq [p]$, leading to the supermodularity of $f(S)$.
	
	Furthermore, observe that $\bar f(S)= {\rm rank} [O(A,C_S);O(A,F)]-{\rm rank} O(A,C_S)$. We can change ${\cal R}(F)$ to ${\cal R} (O(A,F))$, and it is readily seen that the remaining arguments of the above proof are still valid for $\bar f(S)$. 
\end{proof}

\begin{remark}
	From \citep[Theo. 7]{T2016On}, both ${\rm rank} O(A,C_S)$ and ${\rm rank} [O(A,[C_S;F])]$ are submodular on $S\subseteq [p]$. However, there is generally no guarantee that the difference of the two submodular functions is supermodular (or submodular). In this sense, Proposition \ref{sup-numerical} is not a trivial extension of \citep[Theo. 7]{T2016On}.
\end{remark}

%

\begin{proposition} \label{supermodular-graph}
	Define set functions $g(S)\doteq {\rm grank} [O(\bar A,\bar C_S);\bar F]-{\rm grank} O(\bar A,\bar C_S)$ and $\bar g(S)\doteq {\rm grank} [O(\bar A,[\bar C_S;\bar F])]-{\rm grank} O(\bar A,\bar C_S)$, $S\subseteq [p]$. Then, both $g(S)$ and $\bar g(S)$ are supermodular and monotone decreasing.
\end{proposition}

\begin{proof}
	Let the free parameters in $(\bar A,\bar C,\bar F)$ form a vector $\theta$ with dimension $d$. {By regarding $\bar C_S$ as $\bar C$ in the proof of i)$\Rightarrow$ii) in Proposition \ref{characterization-pro}}, for each $S\subseteq [p]$, we get that there is a proper variety of ${\mathbb R}^d$, denoted by ${\mathbb V}_{S}$, such that for all $\theta \in {\mathbb R}^d\backslash {\mathbb V}_S$, ${\rm rank} [O(A,C_S);F]-{\rm rank} O(A,C_S)={\rm grank} [O(\bar A,\bar C_S);\bar F]-{\rm grank} O(\bar A,\bar C_S)=g(S)$, where $(A,C_S,F)$ is obtained by substituting the value of $\theta$ into $(\bar A, \bar C_S, \bar F)$ ({notice that this property holds regardless of whether $(\bar A, \bar C_S, \bar F)$ is SFO or not}).
	Define ${\mathbb V}_0\doteq \bigcup \nolimits_{S\subseteq [p]} {\mathbb V}_S$. Since ${\mathbb V}_0$ is the union of a finite number (at most $2^p$) of proper algebraic varieties, ${\mathbb V}_0$ is still a proper algebraic variety. It turns out that there is a $\hat \theta\in {\mathbb R}^d\backslash {\mathbb V}_0$, denoting the corresponding realization of $(\bar A, \bar C, \bar F)$ as $(A(\hat \theta),C(\hat \theta), F(\hat \theta))$, such that for {\emph{every}} $S\subseteq [p]$, $g(S)={\rm grank} [O(\bar A,\bar C_S);\bar F]-{\rm grank} O(\bar A,\bar C_S)=
	{\rm rank} [O(A(\hat \theta),C(\hat \theta)_S);F(\hat \theta)]-{\rm rank} O(A(\hat \theta),C(\hat \theta)_S)\doteq f(S;\hat \theta)$. In other words, $g(S)$ shares the same property as $f(S;\hat \theta)$ for all $S\subseteq [p]$.
	Since we have proven that $f(S;\hat \theta)$ is supermodular and monotone decreasing for any fixed $\hat \theta \in {\mathbb R}^d$ in Proposition \ref{sup-numerical}, it follows that $g(S)$ is supermodular and monotone decreasing on $S\subseteq [p]$. Similarly, the supermodularity and monotone decreasing of $\bar g(S)$ come from the respective properties of $\bar f(S)$.
\end{proof}

\begin{remark}
The key in the proof of Proposition \ref{supermodular-graph} lies in picking a numerical realization $(A(\hat \theta), C(\hat \theta), F(\hat \theta))$ such that $g(S)=f(S;\hat \theta)$ for {\emph{every}} $S\subseteq [p]$ (similar is for $\bar g(S)$).  It is expected that Proposition \ref{supermodular-graph} can also be proved from a pure graph-theoretic perspective, e.g., using the gammoid structure of linkings (cf. \cite[Chap 2.3]{Murota_Book}). 
\end{remark}

Problem ${\cal P}_1$ (resp. ${\cal P}_2$) can be solved by minimizing $|S|$ subject to $f(S)=0$ (resp. $\bar g(S)=0$). Since $f(S)$ and $\bar g(S)$ are supermodular and monotone decreasing, a simple greedy heuristic can find an approximation solution to the respective problem that is usually close to the optimal solution \cite{krause2014submodular}.  We briefly introduce the procedure of the greedy algorithm. In the following, $h(\cdot)=f(\cdot)$ or $\bar g(\cdot)$ ($g(\cdot)$ is not chosen because it is unclear how to compute it in polynomial time; see Remark \ref{hard}). The greedy algorithm starts with $S_0=\emptyset$. In the $i$th iteration, compute the gain $h(a;S_i)=h(S_i)-h(S_i\cup \{a\})$ for each $a\in [p]\backslash S_i$, select an element $a^*$ with the maximum gain, and add it to $S_i$, i.e.,
$$\begin{array}{c}
a^*=\arg\max\nolimits_{a\in [p]\backslash S_i} h(a;S_i),\\ S_{i+1}=S_i\cup \{a^*\}.
\end{array}$$
The iteration terminates until $h(S_{i})=0$.

We conclude the performance guarantees of the greedy heuristic here, which follow from the known approximation bound for submodular set covering problems, i.e., minimizing $|S|$ subject to
$h(S)=h(V)$, where $h: 2^V\to {\mathbb R}$ is monotone decreasing and supermodular \cite{wolsey1982analysis,krause2014submodular}, thus with proofs omitted. Such an approximation bound is the best one that can be found in polynomial time for general submodular set covering problems~\cite{wolsey1982analysis}.

\begin{theorem}
	Let ${\cal S}_1^*$ and ${\cal S}_2^*$ be respectively the optimal solutions to ${\cal P}_1$ and ${\cal P}_2$. Then, the above-mentioned greedy heuristic based on the set function $f(S)$
	can find a solution ${\cal S}_{f}$ to ${\cal P}_1$ satisfying $$|{\cal S}_{f}|\le (1+\ln ({\rm rank}F)) |{\cal S}_1^*|. $$
	Moreover, the above-mentioned greedy heuristic based on $\bar g(S)$ can find a solution ${\cal S}_{\bar g}$ to ${\cal P}_2$ satisfying $$|{\cal S}_{\bar g}|\le (1+\ln ({\rm grank}O(\bar A, \bar F))) |{\cal S}_2^*|. $$
\end{theorem}

Concerning ${\cal P}_3$, i.e., the problem of selecting the minimal sensors to achieve functional detectability, we have the following result, which follows directly from Proposition \ref{sup-numerical} and the fact that the sum of a finite number of supermodular (resp. monotone decreasing) functions is still supermodular (resp. monotone decreasing) \cite{krause2014submodular}.

\begin{corollary}
	Given a triple $(A,C,F)$, let $(J_i,C_i,F_i)$ be defined as in Section \ref{sec-III}, $i=1,...,k$. Denote ${\rm eig}^+(A)$ as the set of unstable eigenvalues of $A$. Then, the function $f_d(S)\doteq \sum_{\lambda_i\in {\rm eig}^+(A)} ({\rm rank}[O(J_i,C_{iS});F_i]-{\rm rank}O(J_i,C_{iS}))$ is supermodular and monotone decreasing. Consequently, the greedy heuristic based on the set function $f_d(S)$ can find a solution ${\cal S}_{f_d}$ to ${\cal P}_3$ satisfying $|{\cal S}_{f_d}|\le (1+\ln (\sum_{\lambda_i\in {\rm eig}^+(A)}{\rm rank}F_i)) |{\cal S}_3^*|$, where ${\cal S}_3^*$ is the optimal solution to ${\cal P}_3$.
\end{corollary}

\subsection{Modal functional observability based sensor placement} \label{modal_sesnor_placement}
A special case of ${\cal P}_1$ is when there is no prior constraint on the available $C$, that is, $C$ can be arbitrarily designed, termed the unconstrained case. Formally, given a pair $(A,F)$, this problem can be formulated as follows
\begin{equation}\begin{array}{l}
\min \limits_{C\in {\mathbb R}^{p\times n}} p \tag{${\cal P}_4$}\label{A4} \\
{\rm s.t.} \ (A,C,F) \ {\rm functionally \ observable.}
\end{array}\end{equation}
Problem ${\cal P}_4$ asks the minimum number of linear functions of states needed to be measured to estimate $z(t)=Fx(t)$.
When $F=I_n$, i.e., the purpose is to make $(A,C)$ observable, it is known that ${\cal P}_4$ has a closed-form solution, being the maximum geometric multiplicity of eigenvalues of matrix $A$ \cite{Tong2017Minimal}. In what follows, based on the proposed modal functional observability, we generalize this result to ${\cal P}_4$ on the class of diagonalizable systems.  Notice that diagonalizable systems commonly arise in practical settings, such as systems whose state matrices are symmetric or have simple spectra \cite{zhang2018observability}. 

As $(J_i,C_i,F_i)$ in Definition \ref{modal_functional_ob} can be complex valued, in order to construct a real output matrix $C$, the {\emph{real Jordan normal form}} of diagonalizable matrices is introduced; see Section \ref{sec-III} and \citep[Chap. 3.4]{Horn2013Matrix} for the general case.  Suppose $A\in {\mathbb R}^{n\times n}$ has $2k_c$ distinct non-real eigenvalues and $k_r$ distinct real eigenvalues, of which the $k_c$ pairs of complex conjugate eigenvalues are denoted by $\lambda_i=a_i+b_i{\bf i}$ and $\lambda_{k+1-i}=\lambda_i^*=a_i-b_i{\bf i}$, $a_i\in {\mathbb R}$, $b_i>0$, $i=1,...,k_c$, and $\lambda_{k_c+1},...,\lambda_{k_c+k_r}$ are real eigenvalues (accordingly, $k=2k_c+k_r$).
When $A$ is diagonalizable, in its real Jordan normal form, the diagonal block associated with a non-real $\lambda_i=a_i+b_i{\bf i}$ (also $\lambda_i^*$) with algebraic multiplicity $d_i$
is given by $\tilde J_{i}={\bf diag}\{D_i,...,D_i\}\in {\mathbb R}^{2d_i\times 2d_i}$ with $D_i={\tiny \left[\begin{array}{cc}
 	a_i & b_i \\
 	-b_i & a_i
 	\end{array} \right]}$. Let $\Phi={\tiny \left[\begin{array}{cc}
	-{\bf i} & -{\bf i}\\
	1 &  -1
	\end{array}\right]}$. As $D_i=\Phi{\tiny \left[\begin{array}{cc}
	\lambda_i & 0\\
	0 &  \lambda_i^*
	\end{array}\right]}\Phi^{-1}$, the complex Jordan normal form $T^{-1} A T= J$ and the real one $\tilde T^{-1} A \tilde T= J_r$ can be related as follows. Let $t_{i1},...,t_{i{d_i}}$ be $d_i$ linearly independent eigenvectors associated with a {\emph{non-real}} eigenvalue $\lambda_i$ (then $t^*_{i1},...,t^*_{i{d_i}}$ are the corresponding eigenvectors associated with $\lambda_i^*$), comprising sub-columns of $T$, and let $[\tilde t_{i1},...,\tilde t_{i2d_i}]$ be the columns of $\tilde T$ corresponding to $\tilde J_i$. Then, $[\tilde t_{i1},...,\tilde t_{i2d_i}]=[t_{i1},t_{i1}^{*},...,t_{id_i},t_{id_i}^{*}]\Phi^{-1}_{d_i}$, where $\Phi_{d_i}={\bf diag}\{\Phi,...,\Phi\}\in {\mathbb C}^{2d_i\times 2d_i}$. {Meanwhile, for a {\emph{real}} eigenvalue $\lambda_i$, the corresponding $t_{i1},...,t_{id_i}$ and $\tilde t_{i1},...,\tilde t_{id_i}$ coincide.} Consequently, there exists an $n\times n$ permutation matrix $\Gamma$ such that $\tilde T=T\Gamma\tilde \Phi^{-1}$, with $\tilde \Phi\doteq \{\Phi_{d_1},...,\Phi_{d_{k_c}},I_{d_{k_c+1}},...,I_{d_{k_c+k_r}}\}$.
%


\begin{theorem} \label{minimal-sensor-analytical-theorem}
	Given a pair $(A,F)$, suppose $A$ is diagonalizable. Let $(J_i,F_i)$ be defined as in Section \ref{sec-III}, $i=1,...,k$. Then, the minimum number of sensors making $(A,C,F)$ functionally observable, denoted as $p^*$, is $$p^*=\max_{1\le i \le k} {\rm rank} F_i,$$and, there is a deterministic procedure that can determine such a $C\in {\mathbb R}^{p^*\times n}$ in polynomial time.
\end{theorem}

\begin{proof} Suppose ${\hat C}$ is a matrix such that $(A,{\hat C}, F)$ is functionally observable. By Theorem \ref{theorem-func-observe}, ${\rm rank}[O(J_i,{\hat C}_i);F_i]={\rm rank}O(J_i,{\hat C}_i)$ should hold for each $i\in [k]$, where ${\hat C}_i={\hat C}T_i$. Since $A$ is diagonalizable, this is equivalent to (c.f. Lemma \ref{diagonal-modal})
	$${\rm rank} [{\hat C}_i;F_i]={\rm rank} {\hat C}_i,$$
	which requires ${\rm row}({\hat C}_i)\ge {\rm rank}F_i$. As a result, ${\rm row}({\hat C})\ge {\rm rank}F_i$ for each $i\in [k]$, meaning $p^*$ lower bounds ${\rm row}({\hat C})$ for any ${\hat C}$ making $(A,{\hat C}, F)$ functionally observable.
	

{We shall give a procedure to construct a {$C\in {\mathbb R}^{p^*\times n}$ and then prove that $(A,C, F)$ is functionally observable}.} Suppose $J_r=\tilde T^{-1}A\tilde T={\bf diag}\{\tilde J_{1},...,\tilde J_{k_c},...,\tilde J_{k_c+k_r}\}$ is the real Jordan normal form of $A$ as described above. Let $\tilde F= F\tilde T$, and partition $\tilde F=[\tilde F_1,...,\tilde F_{k_c+k_r}]$, in which $\tilde F_i$ has the same number of columns as $\tilde J_{i}$, $i=1,...,k_c+k_r$. For  $i\in [k_c+k_r]$, let $\tilde C'_i$ consist of $\tilde d_i$ linearly independent rows of $\tilde F_i$ such that {\begin{equation}\label{real} {\rm rank}\left[\begin{array}{c}
		\tilde J_i-\lambda_iI \\
		\tilde F_i \\
		\tilde C_i'
		\end{array} \right]={\rm rank}\left[\begin{array}{c}
		\tilde J_i-\lambda_iI \\
		\tilde C_i'
		\end{array} \right],\end{equation}}with $\tilde d_i={\rm rank}[\tilde J_i-\lambda_iI;\tilde F_i]-{\rm rank}(\tilde J_i-\lambda_iI)$.  Such a $\tilde C'_i$ can be found by iteratively choosing one row from $\tilde F_i$ to increase the rank of $[\tilde J_i-\lambda_i I;\tilde C'_i]$ by one per iteration until it reaches ${\rm rank}[\tilde J_i-\lambda_i I;\tilde F_i]$,  Moreover, let $\tilde C_i=[\tilde C_i';0_{(p^*-\tilde d_i)\times 2d_i}]$ for $i\in [k_c]$ and
	$\tilde C_i=[\tilde C'_{i};0_{(p^*-\tilde d_{i})\times d_{i}}]$ for $i\in [k_c+k_r]\backslash [k_c]$, i.e.,  adding zero matrices of appropriate dimensions to $\tilde C_i'$ to make them have the same number of rows (why $\tilde d_i\le p^*$ is explained later).  Construct ${C}\in {\mathbb R}^{p^*\times n}$ as
	$${C}=[\tilde C_1,...,\tilde C_{k_r+k_c}]\tilde T^{-1}.$$
	Since the real Jordan normal form can be obtained in polynomial time \cite{Horn2013Matrix}, ${C}$ can be determined in polynomial time.
	
We are to prove that $(A,{C},F)$ is functionally observable. {By definition, for each real eigenvalue $\lambda_i$ of $A$, its real Jordan block and complex one coincide, leading to $\tilde J_i=J_i$ and $\tilde F_i=F_i$. Moreover, notice that $CT=[\tilde C_1,...,\tilde C_{k_r+k_c}]\tilde T^{-1}T=[\tilde C_1,...,\tilde C_{k_r+k_c}]\tilde \Phi \Gamma^{-1}$, resulting in $C_i\doteq CT_i=\tilde C_i$, where $\tilde T=T\Gamma\tilde \Phi^{-1}$ has been used. The construction (\ref{real}) then yields that each real eigenvalue of $A$ is modal functionally observable. Therefore, from Theorem \ref{theorem-func-observe}, it remains to prove that each non-real eigenvalue of $A$ is modal functionally observable.}

	For each $i\in [k_c]$, observe that
	$$\begin{array}{c}
	{\rm rank}\left[\begin{array}{c}
	\tilde J_i-\lambda_iI \\
	\tilde F_i \\
	\tilde C_i
	\end{array} \right]={\rm rank}\left[\begin{array}{cc}
	\Phi_{d_i}^{-1} & 0 \\
	0 & I
	\end{array} \right]\left[\begin{array}{c}
	\tilde J_i-\lambda_iI \\
	\tilde F_i \\
	\tilde C_i
	\end{array} \right]\Phi_{d_i} \\
	={\rm rank}\left[\begin{array}{cc}
	J_i-\lambda_iI & 0 \\
	0 & J_{k+1-i}-\lambda_iI \\
	F_i & F_{k+1-i} \\
	C_i & C_{k+1-i}
	\end{array}
	\right]={\rm rank}\left[\begin{array}{c}
	F_i\\
	C_i
	\end{array}
	\right]+d_i,
	\end{array}$$where the second equality is the result of some column permutations.
	Similarly, by looking at the first two row blocks and the first and third row blocks of the left most matrix in the above relation,  we have ${\rm rank}[\tilde J_i-\lambda_iI;\tilde F_i]={\rm rank}F_i+d_i$ and ${\rm rank}[\tilde J_i-\lambda_iI;\tilde C_i]={\rm rank}C_i+d_i$. Note that ${\rm rank}(\tilde J_i-\lambda_i I)=d_i$,  which leads to $\tilde d_i={\rm rank}[\tilde J_i-\lambda_iI;\tilde F_i]-{\rm rank}(\tilde J_i-\lambda_iI)={\rm rank} F_i\le p^*$. Since ${\rm rank}[\tilde J_i-\lambda_iI;\tilde F_i;\tilde C_i]={\rm rank}[\tilde J_i-\lambda_iI;\tilde F_i]={\rm rank}[\tilde J_i-\lambda_iI;\tilde C_i]$ by the construction of $\tilde C_i'$ in (\ref{real}), we have ${\rm rank}[F_i;C_i]={\rm rank}C_i$, meaning that $\lambda_i$ is modal functionally observable. 	
	Moreover, as $C$ is real, it follows that $F_{k+1-i}=F_i^*$ and $C_{k+1-i}=C_i^*$ (notice that the eigenvectors associated with $\lambda_i$ and with $\lambda_{k+1-i}=\lambda_i^*$ are complex conjugate). Because ${\rm rank}C_i^*={\rm rank}C_i$ and ${\rm rank}[F_i^*;C_i^*]={\rm rank}[F_i;C_i]$ (\citep[Lem 1]{zhang2022real}), the modal functional observability of $\lambda_{k+1-i}$ follows from the same property of $\lambda_i$, $\forall i\in [k_c]$.
\end{proof}

\begin{remark} By considering only unstable modes, Theorem \ref{minimal-sensor-analytical-theorem} can be easily extended to the functional detectability case. When $A$ is not necessarily diagonalizable, let $p_i^*$ be the minimum ${\rm row}(C_i)$ such that (\ref{modal_functional}) holds. It follows from Theorem \ref{theorem-func-observe} that the minimum ${\rm row}(C)$ making $(A,C,F)$ functionally observable then equals ${\max}_{1\le i\le k}\ p_i^*$. However, as explained in Remark \ref{challenge-eigenspace}, determining $p_i^*$ in the non-diagonalizable case is more complicated, which is left for future exploration.
\end{remark}


\begin{remark}It can be seen that when $F=I_d$, then $p^*= \max_{i\in [k]} d_i$, which coincides with the maximum geometric multiplicity solution for the unconstrained minimal observability problem \cite{zhang2018observability}. On the other hand, if $(A,C)$ is observable for some $C\in {\mathbb R}^{p^*\times n}$, then an unstructured random $C'\in {\mathbb R}^{p^*\times n}$ (meaning that $C'$ does not have any prescribed structure) will make $(A,C')$ observable with probability one.\footnote{{To see this, consider $C'$ as a structured matrix full of $*$ entries. As $(A,C)$ is observable, from the PBH test, for each $i\in [k]$, upon letting matrix $X_i\in {\mathbb C}^{n\times g_i}$ consist of a set of linearly independent vectors spanning the eigenspace associated with $\lambda_i$, where $g_i$ is the geometric multiplicity of $\lambda_i$, we obtain that $CX_i$ has full column rank. By regarding $C$ as a realization of $C'$, we get that there is a $g_i\times g_i$ submatrix in $C'X_i$, whose determinant is not identically zero. Consequently, there is a proper variety in ${\mathbb R}^{p^*\times n}$, given by ${\mathbb V}_i$, such that for each realization of $C'$ in ${\mathbb R}^{p^*\times n}\backslash {\mathbb V}_i$, the aforementioned determinant is not zero, implying that the corresponding $C'X_i$ has full column rank. It follows that for each realization in ${\mathbb R}^{p^*\times n}\backslash (\bigcup_{i\in [k]} {\mathbb V}_i)$, the corresponding $C'X_i$ has full column rank, $\forall i\in [k]$, leading to the observability of $(A,C')$. As $\bigcup_{i\in [k]} {\mathbb V}_i$ has zero measure in ${\mathbb R}^{p^*\times n}$, the claim is immediate.}} However, things are different for functional observability, as the matrix $C\in {\mathbb R}^{p^*\times n}$ making $(A,C,F)$ functionally observable may fall in some hyperplane in ${\mathbb R}^{p^*\times n}$.
\end{remark}

We end this section by providing an example to illustrate the procedure mentioned in the proof of Theorem \ref{minimal-sensor-analytical-theorem}.
\begin{example}
	Consider a pair $(A,F)$ as
	$$\begin{array}{c}A=\left[\begin{array}{cccccccc}
	0& 4& 3&-2&-3& 0&-4& 1\\
	-1&-1&-1& 1& 0& 3& 1& 4\\
	2&10& 1&-3&-3& 3&-3& 5\\
	-1& 9& 1&-2&-3& 9&-1&13\\
	0& 0& 0& 0& 1& 1& 0& 0\\
	0& 0& 0& 0&-1& 1& 0& 0\\
	0& 0& 0& 0& 0& 0& 1& 1\\
	0& 0& 0& 0& 0& 0&-1& 1
	\end{array}\right]\\
	F=[I_4,0_{4\times 4}].
	\end{array}$$
	Simple calculations show that $A$ has two pairs of complex conjugate eigenvalues, being $1\pm {\bf i}$ (with algebraic multiplicity $d_1=d_4=3$) and $-2\pm {\bf i}$ (with algebraic multiplicity $d_2=d_3=1$), respectively. There exists a real invertible matrix $\tilde T\in {\mathbb R}^{8\times 8}$ that transforms $A$ to its real Jordan normal form $J_r$, with
	$$\begin{array}{c}J_r=\tilde T^{-1}A\tilde T={\bf diag}\{D_1,D_1,D_1,D_2\}\\
	\tilde F \doteq  F\tilde T=[\tilde F_1, \tilde F_2]=\left[\begin{array}{cccccc|cc}
	1&3&2&2&1&0&1&0\\
	0&1&2&1&0&1&0&0\\
	1&4&4&3&1&1&0&1\\
	1&6&8&5&1&3&1&1
	\end{array}\right] \end{array},$$
where$$\begin{array}{c}\tilde  T=\left[\begin{array}{c} \tilde F \\ \left[I_4,0_{4\times 4}\right] \end{array} \right], D_1=\left[\begin{array}{cc} 1 & 1\\ -1 & 1 \end{array} \right], D_2=\left[\begin{array}{cc} -2 & 1\\ -1 & -2 \end{array} \right].
	\end{array} $$
As a result, $\tilde d_1=2, \tilde d_2=1$, $p^*=\max \{\tilde d_1,\tilde d_2\}=2$. By the procedure mentioned in the proof of Theorem \ref{minimal-sensor-analytical-theorem}, we can choose the first two rows from $\tilde F_1$ and the last row from $\tilde F_2$ to form a matrix $\tilde C=[\tilde C_1,\tilde C_2]$ as
	$$\tilde C=\left[\begin{array}{cccccc|cc}   1&3&2&2&1&0&1&1\\
	0&1&2&1&0&1&0&0  \end{array}\right].$$Note that the choice is not unique. The required output matrix with the minimum number $p^*=2$ of rows is then
	$$C=\tilde C\tilde T^{-1}=\left[\begin{array}{cccccc|cc}  0&-3& 0& 1& 0& 0& 0& 0 \\
	0& 1& 0& 0& 0& 0& 0& 0  \end{array}\right]. $$
	It can be verified that ${\rm rank}O(A,C)=6={\rm rank}[O(A,C);F]$, implying $(A,C,F)$ is indeed functionally observable. Meanwhile, adding unstructured random perturbations to $C$ to obtain $C'$ makes  ${\rm rank}O(A,C')=6\ne {\rm rank}[O(A,C');F]=8$ with overwhelming probability. This indicates the output matrix $C$ with the minimum number of rows making $(A,C,F)$ functionally observable may fall in some hyperplane of the space ${\mathbb R}^{2\times 8}$.
	
	On the other hand, we can also use the greedy algorithm to select dedicated sensors for achieving functional observability. It happens that deploying $2$ dedicated sensors at the first two state variables is sufficient for this goal.
\end{example}

\section{Conclusions}
In this paper, motivated by the fact that the existing PBH-like conditions for functional observability and detectability are only valid for diagonalizable sytems, we developed novel characterizations for functional observability, functional detectability, and SFO {that are valid for general systems}. We first introduced a new notion {\emph{modal functional observability} to establish new necessary and sufficient conditions for functional observability and detectability. Then, we rigorously redefined SFO from a generic perspective,  highlighting its distinctions from the conventional structural observability. A complete graph-theoretic characterization for SFO is further put forward. Building upon these results, we proved that the minimal sensor placement problems for functional observability and SFO are NP-hard. Nevertheless, we devised efficient algorithms that leverage supermodular set functions to find near-optimal solutions. Additionally, we offered a closed-form solution based on modal functional observability and the real Jordan normal form for deploying the minimum number of sensors to achieve functional observability for diagonalizable systems. {Exploring the generalization of this result to non-diagonalizable systems and structured systems is left for future research.}

	\bibliographystyle{elsarticle-num}
	{\footnotesize
		\bibliography{yuanz3}
	}


\begin{IEEEbiography}[{\includegraphics[width=0.8in,height=1in,clip,keepaspectratio]{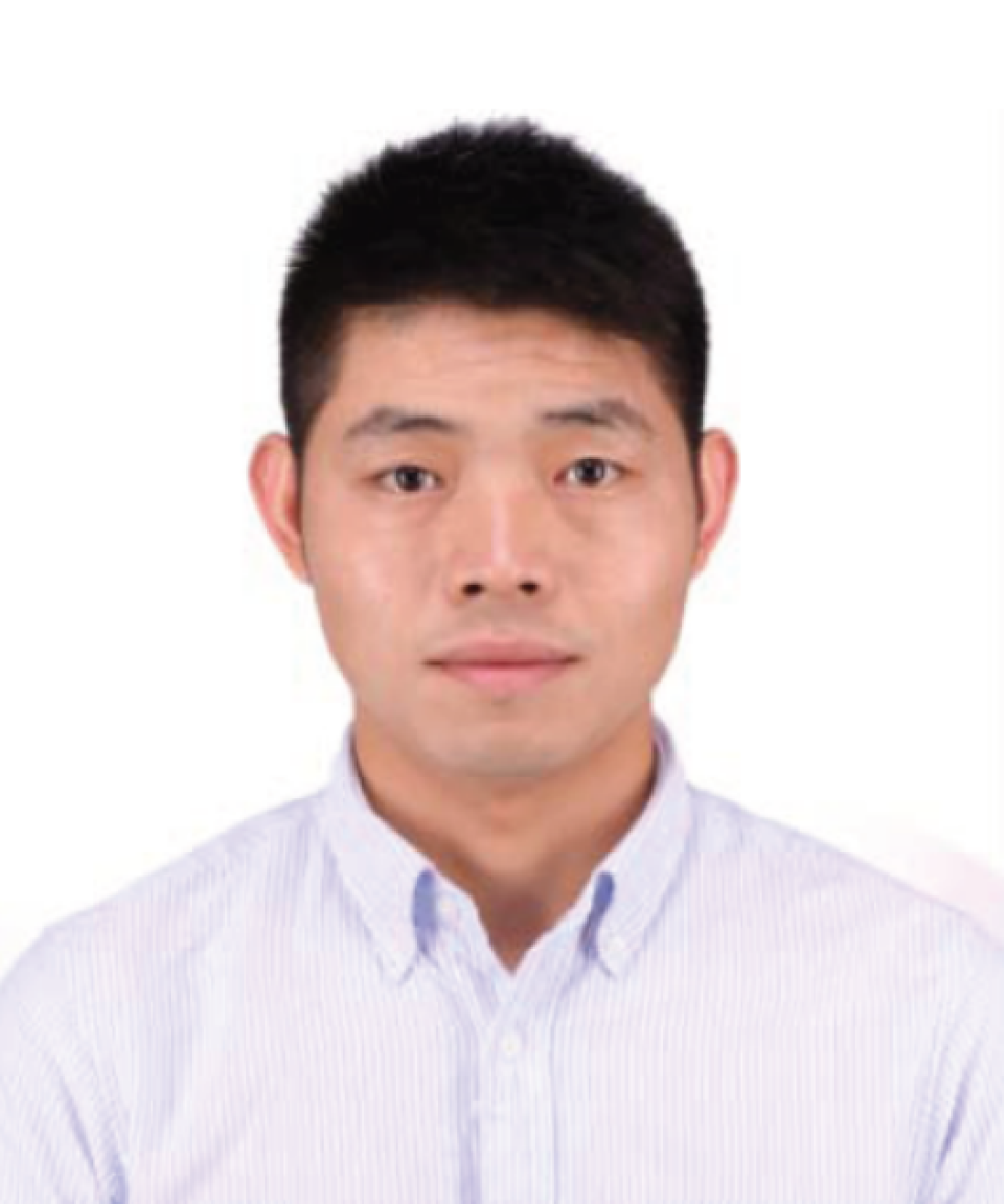}}]{{\bf Yuan Zhang}}
\; (Member, IEEE) received the B.S. degree in control science and
			engineering from Tongji
			University, Shanghai, China in 2014, and the Ph.D
			degree in control science and
			engineering from Tsinghua University, Beijing, China in
			2019. Since then, he
			has worked in Beijing Institute of Technology, Beijing,
			China, first as a
			postdoctoral fellow, and currently an associate
			professor. He was a visiting research fellow at University of Western Australia, Australia in 2024.

He received the Excellent Doctoral Dissertation Award from Tsinghua University, and was selected in the China Postdoctoral Innovative Talent
            Support Program, both in 2020. His research
			interests include analysis, control and optimization of network
			systems and data-driven analysis.
\end{IEEEbiography}

\begin{IEEEbiography}[{\includegraphics[width=0.95in,height=1in,clip,keepaspectratio]{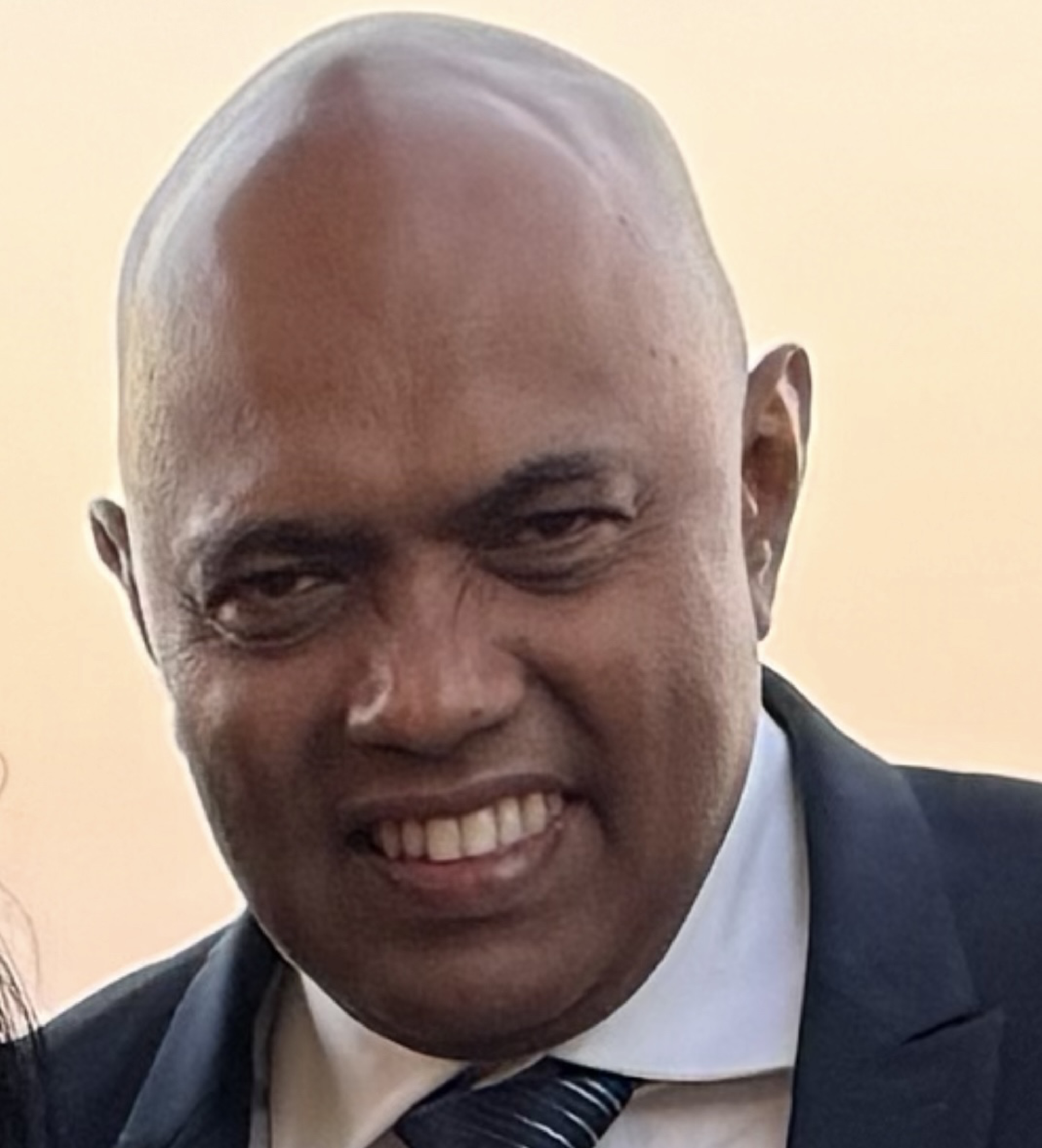}}]{{\bf {\bf Tyrone Fernando}}}
\; (Senior Member, IEEE) earned his Bachelor of Electrical Engineering (Hons.) and Ph.D. from the University of Melbourne, Australia, in 1990 and 1996, respectively. In 1996, he joined the Department of Electrical, Electronic, and Computer Engineering at the University of Western Australia (UWA) in Crawley, WA, where he currently serves as a professor. From 2008 to 2010, he held the roles of Associate Head and Deputy Head of the department. Prof. Fernando is presently the Head of the Power and Clean Energy Research Group at UWA. Since 2021, he has also provided ongoing consultancy to Western Power, focusing on the management of distributed energy resources within the electric grid. His research interests encompass control systems, observer design, and applications in power systems. In 2018, he was recognized as the Outstanding WA PES/PELS Engineer.

Prof. Fernando has served as an associate editor for several scholarly journals and has received multiple teaching awards from UWA for his contributions to control systems and power systems education.
%
\end{IEEEbiography}

\begin{IEEEbiography}[{\includegraphics[width=1.0in,height=1.1in,clip,keepaspectratio]{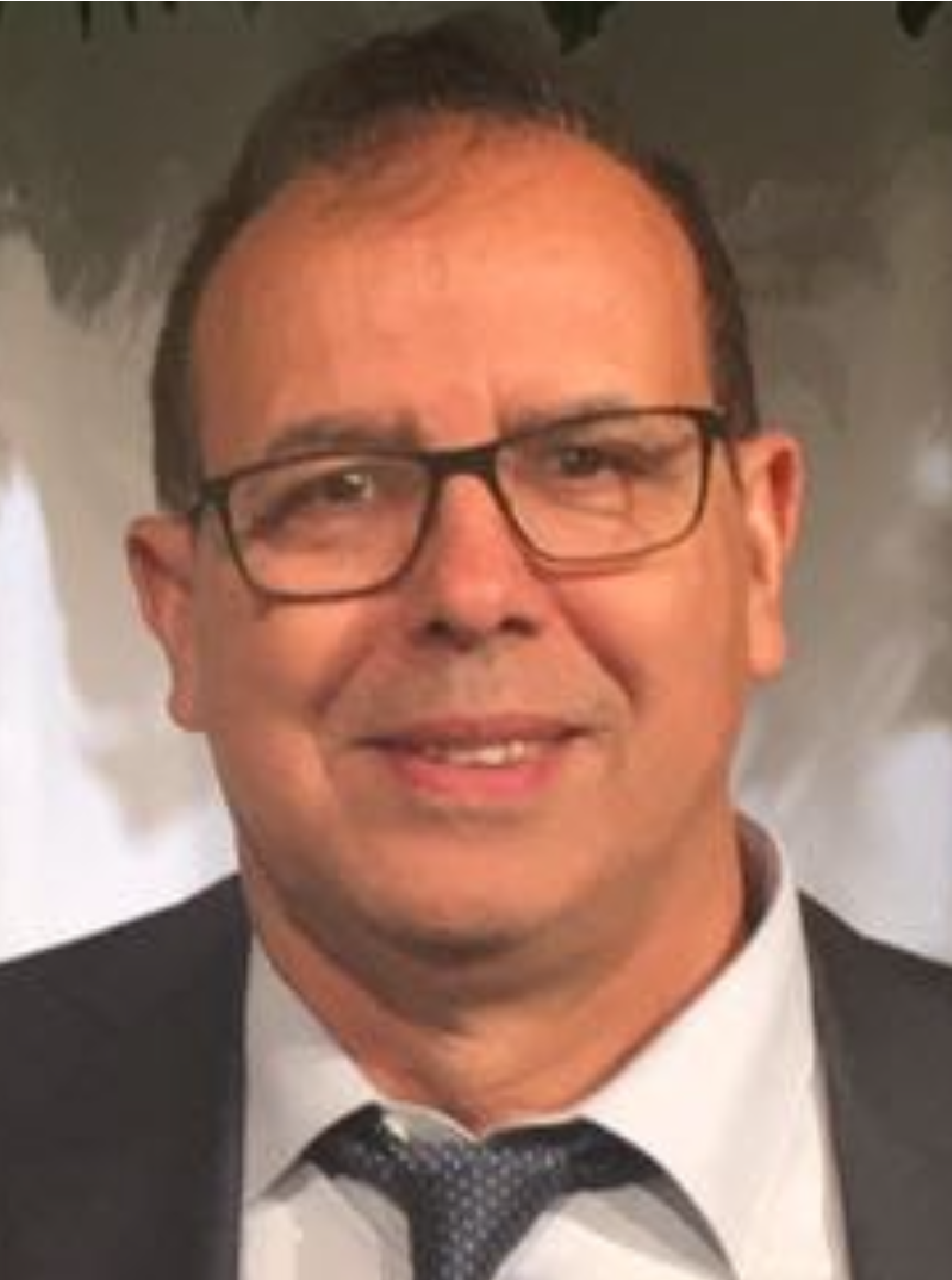}}]{{\bf {\bf Mohamed Darouach}}}
 received his graduate degree in electrical engineering from Ecole Mohammadia d'Ingenieurs, Rabat, Morocco, in 1978, the Doctor Engineer in automatic control and Doctor of Sciences degrees in Physics from Nancy University, Nancy, France, in 1983 and 1986, respectively, and the Honoris Causa degree from the Technical University of IASI, Iasi, Romania.

From 1978 to 1986, he was an Associate Professor and a Professor of automatic control at Ecole Hassania des Travaux Publics, Casablanca, Morocco. From 1987 to 2024 he was a Professor at Universit\'e de Lorraine, France. Since 2024, he has been Distinguished Professor Emeritus at Universit\'e de Lorraine, France. He was a Vice Director of the Research Center in Automatic Control of Nancy (CRAN UMR 7039, Universite de Lorraine, CNRS)
from 2005 to 2013. Since 2010, he is a member of the scientific council of Luxembourg University. From 2013 to 2018 he was a Vice Director of the University Institute of Technology of Longwy (Universit\'e de Lorraine). He held invited positions at University of Alberta, Edmonton, Canada, and University of Western Australia, Perth, Australia. Since 2017, he has been a Robert and Maude Gledden Senior Visiting Fellow, administrated by the Institute of Advanced Studies at UWA. In 2019, 2022 and 2023 he also held a Forrest Visiting Fellow, UWA. His research interests include theoretical control, and observers design and control of large-scale systems with applications.

\end{IEEEbiography}


\end{document}